# Optimizing Performance and Satisfaction in Matching and Movement Tasks in Virtual Reality with Interventions Using the Data Visualization Literacy Framework

**Andreas Bueckle[1]*, Kilian Buehling[2], Patrick C. Shih[3], Katy Börner[1]**

[1] Department of Intelligent Systems Engineering, Luddy School of Informatics, Computing, and Engineering, Indiana University, Bloomington, IN, USA

[2] Research Group Knowledge and Technology Transfer, Fakultät Wirtschaftswissenschaften, Technische Universität Dresden, Germany

[3] Department of Informatics, Luddy School of Informatics, Computing, and Engineering, Indiana University, Bloomington, IN, USA

*** Correspondence:**
Corresponding Author
abueckle@iu.edu

**Abstract**

Virtual reality (VR) has seen increased use for training and instruction. Designers can enable VR users to gain insights into their own performance by visualizing telemetry data from their actions in VR. Our ability to detect patterns and trends visually suggests the use of data visualization as a tool for users to identify strategies for improved performance. Typical tasks in VR training scenarios are manipulation of 3D objects (e.g., for learning how to maintain a jet engine) and navigation (e.g., to learn the geography of a building or landscape before traveling on-site). In this paper, we present the results of the RUI VR (84 subjects) and Luddy VR studies (68 subjects), where participants were divided into experiment and control cohorts. All subjects performed a series of tasks: 44 cube-matching tasks in RUI VR , and 48 navigation tasks through a virtual building in Luddy VR (all divided into two sets). All Luddy VR subjects used VR gear; RUI VR subjects were divided across three setups: 2D Desktop (with laptop and mouse), VR Tabletop (in VR, sitting at a table), and VR Standup (in VR, standing). In an intervention called "Reflective phase," the experiment cohorts were presented with data visualizations, designed with the Data Visualization Literacy Framework (DVL-FW), of the data they generated during the first set of tasks before continuing to the second part of the study. For Luddy VR, we found that experiment users had significantly faster completion times in their second trial ($\boldsymbol{p} = \mathbf{0.014}$) while scoring higher in a mid-questionnaire about the virtual building ($\boldsymbol{p} = \mathbf{0.009}$). For RUI VR, we found no significant differences for completion time and accuracy between the two cohorts in the VR setups; however, 2D Desktop subjects in the experiment cohort had significantly higher rotation accuracy as well as satisfaction ($\boldsymbol{p_{rotation}} = \mathbf{0.031}$, $\boldsymbol{p_{satisfaction}} = \mathbf{0.040}$). We conclude with suggestions for adjustments to the Reflective phase to boost user performance before generalizing our findings to performance improvement in VR with data visualizations.

## 1 Introduction

Due to decreasing cost and an increasing amount of hardware choice, VR has become a popular entertainment tool. Recent devices such as the Oculus Quest 2 offer VR at low prices and without the need for a strong PC or laptop. Further, there is an increasing market for coaching and training employees for retail, maintenance, and administrative jobs. As upskilling the workforce becomes an ever-larger challenge nationally and internationally, immersive tools have become a viable option of teaching employees in many industries to assemble jet engines, manufacture specialized automobile parts, assemble electrical circuits, and learn the layout of a building to give physical tours or otherwise facilitate events via VR training applications. Given that VR applications are data-rich, information visualization becomes a viable tool to allow trainees to reflect upon and improve their own performance.

Depending on the hardware and the needs of the application, users of VR equipment generate position and rotation data at a rate of up to 120 Hz, and every button press can be logged and associated with a time stamp via telemetry. In addition to these physical variables, additional data can be derived via computation at runtime or in later analysis, allowing designers and researchers to measure a user's performance and behavior when completing tasks such as arranging objects or navigating spaces. The novelty of VR, while demonstrably exciting and invoking a feeling of presence (Batch et al., 2019), brings with it challenges due to its unfamiliarity to many users. With the basis of training being repetition and improvement over time, methods to assess and improve one's performance are necessary. The visual primacy of VR, along with the availability of user data, suggests data visualization as a good tool to allow users to gain insights into their own performance.

## 1.1 Overview

In this paper, we describe two user studies where we developed interventions to improve VR performance for manipulation ("**RUI VR**") and navigation tasks ("**Luddy VR**"). These VR visualizations were developed using the Data Visualization Literacy Framework or DVL-FW (Börner and Polley, 2014; Börner, 2015; Börner et al., 2019), a theoretical toolset to interpret, construct, and teach data visualizations. The DVL-FW comes with a series of seven typologies to categorize, among others, visualization types (such as graphs and maps), visual encodings via graphic symbols (such as points, lines, volumes) and graphic variables (such as color hue and size), and interactions with data (such as filter as well as link and brush). We use the DVL-FW to describe the data visualization interventions with an abstracted terminology that expresses both traditional, 2D data visualizations (like bar graphs and line graphs) and advanced VR visualizations. Of special interest is the implementation of four interaction types (**filter, navigate, animate/replay, link and brush**): We enabled the subjects to filter their data by time stamp or graphic symbol (RUI VR) and task number (both studies); users could navigate freely around their data, which was displayed in its original spatial context on a 1:1 scale (RUI VR) and minimized (Luddy VR); and it was possible to play back the data by time stamp in different speeds via a time slider (RUI VR); and subjects could select bars in a bar graph and then apply filters correspondingly to view only specific tasks based on their completion time. The goal of these studies was to test whether significant differences in performance and satisfaction were measurable between the control and experiment cohorts and to determine the effects between behavior in the intervention and performance in the subsequent set of tasks.

For the **RUI VR study**, we found **no significant differences** between the two VR cohorts for mean position accuracy, rotation accuracy, and completion time. However, we found that the experiment cohort for the 2D Desktop setup achieved **higher rotation accuracy and satisfaction** (Mann-Whitney-U-Test (Mann and Whitney, 1947), $p_{rotation}$ **= 0.031,** $p_{satisfaction}$ **= 0.04**). Likewise, the experiment cohort for **VR Standup** reported significantly higher satisfaction scores (Mann-Whitney-U-Test, $\boldsymbol{p}$ **= 0.016**).

In **Luddy VR**, on the other hand, subjects in the experiment cohort achieved **significantly faster completion times** in their second trial (Welch's t-test, $\boldsymbol{p}$ **= 0.014**) while also **scoring higher in a mid-questionnaire** about the topology of the virtual building that they navigated through than their control cohort counterparts (Welch's t-test, $\boldsymbol{p}$ **= 0.008734**), prompting us to conclude that the Reflective phase allowed our users to **derive better strategies** for completing their navigation tasks.

## 1.2 Related Work

There is extensive prior work on the use of VR and other immersive technologies along the reality-virtuality continuum (Milgram and Kishino, 1994; Skarbez et al., 2021) for visualizing scientific data. Examples include assessing risk in mining with seismic data (Kaiser et al., 2005), protein-docking (Anderson and Weng, 1999), astronomy (Djorgovski et al., 2013; Donalek et al., 2014), computer-aided design (Smets et al., 1993), geographical information science (Huang et al., 2001), and health care (Ibrahim and Money, 2019). Bryson (1996) pointed at the affordances that VR offers for interaction with complex phenomena and their representations in data: "We want to create the effect of interacting with things, not with pictures of things" (p. 63).

### 1.2.1 Immersive Analytics

More recently, the emerging field of Immersive Analytics (Chandler et al., 2015; Bach et al., 2016; Simpson et al., 2016; Dwyer et al., 2018; Marriott et al., 2018) has put a focus on information visualization of abstract datasets, such as network connectivity (Cordeil et al., 2017b) and economic data (Batch et al., 2019). Software development kits such as IATK (Cordeil et al., 2019) and ImAxes (Cordeil et al., 2017a) enable researchers and designers to craft visualization of economic, health, and other non-spatial data. These allow for the creation of what Bowman et al. (2003) called "information-rich environments" where the virtual objects are representations of non-physical entities.

Although VR lends itself to visualizing both spatially explicit and abstract data, to the best of our knowledge, the use of data visualization to empower users to optimize performance in VR has not been studied to the same degree. For example, while there is prior work investigating qualitative user feedback on the feeling of presence for types of navigation in VR (Slater et al., 1995a), on testing efficiency vs. presence (Slater et al., 1995b), or a combination of human factors (Usoh et al., 1999), none of these studies tested user improvement after investigating their own data. Likewise, VR studies involving cube-matching tasks have focused on testing interaction paradigms without giving users the ability to gain insight into their own performance (Mendes et al., 2017).

### 1.2.2 Data Visualization Literacy

Given the increasing use of VR for workplace training, there is a growing need to enable users to optimize their performance. A viable tool to achieve this goal is data visualization, owing to our ability to detect visual patterns and trends. Many attempts have been made to formalize how to construct, interpret, and teach the growing "visualization zoo" (Heer et al., 2010), with a focus on tasks (Amar and Stasko, 2004; Brehmer and Munzner, 2013), interactions (Yi et al., 2007; Roth, 2013), and graphic symbols and variables (Cleveland and McGill, 1984; Heer and Bostock, 2010). The Data Visualization Literacy Framework or DVL-FW (Börner and Polley, 2014; Börner, 2015; Börner et al., 2019) has been developed on the basis on many of the aforementioned works to "define, teach, and assess [data visualization literacy]". As data becomes increasingly prevalent in our everyday lives, skills relating to the understanding of trends, patterns, and structures of temporal, geospatial, topical, and network data are increasingly important for professional and personal decision-making. Unlike other literacy types such as numeracy (Reyna et al., 2009; OECD, 2013a), textual literacy (OECD, 2013b), or visual literacy (Fransecky and Debes, 1972; Ausburn and Ausburn, 1978; Avgerinou, 2007; Hattwig et al., 2013), data visualization literacy has seen formal assessment attempts only very recently. Boy et al. (2014) proposed two tests for visualization literacy with line graphs using item response theory, successfully validating their model with a user study involving 40 subjects on Amazon Mechanical Turk (MTurk, https://www.mturk.com/). With similar methodological goals, Lee et al. (2017) employ test development in psychology and education to develop the Visual Literacy Assessment Test (VLAT), designed to measure how non-expert users interpret data visualizations. They validate their test, consisting of 12 data visualizations, 53 multiple-choice items, and eight visualization tasks, using input from five domain experts in data visualization and a user study with 191 MTurk subjects to show a high reliability. Making sense of unfamiliar visualizations continues to pose a challenge, especially for novices (Börner et al., 2016; Lee et al., 2016). It is to be expected that issues related to sense-making would translate into the realm of VR.

### 1.2.3 Data Visualization in VR

While data visualizations in 2D space have been extensively studied (and effectively are the standard), data visualizations in 3D (and, by extension, VR) have been met with caution in the literature, with authors warning of “unjustified 3D” (Munzner, 2014) and calling to “[a]void 3-D displays of quantitative data” (Few, 2012). At the same time, there is a growing body of literature arguing for the benefits of 3D or VR for data visualization (as opposed to the aforementioned visualization of scientific data), focused on properties such as presence (Batch et al., 2019), embodiment (Jacob et al., 2008), and involvement (Rosenbaum et al., 2011), as well as specific domains, e.g., aerospace engineering (García-Hernández et al., 2016). Several studies have also been conducted to compare 2D and VR implementations for visualizations. Raja et al. (2004) performed a pilot study with four conditions and four subjects using the CAVE (Cruz-Neira et al., 1993), and found that the most immersive one (content on all four walls, with head-tracking) yielded the best results in terms of the users’ ability to view datasets and complete tasks. Millais et al. (2018) presented a 3D scatter graph and a 3D parallel coordinate plot to 16 subjects, half of them using a 2D screen, half using VR gear, in a think-aloud session. They found no significant difference in workload between the 2D and VR data exploration but determined that data exploration felt more successful and satisfying for VR users. Similarly, these users also reported that the physical demand was higher.

## 2 Materials and Methods

Both the studies presented here followed the same general study design, see Figure 1.

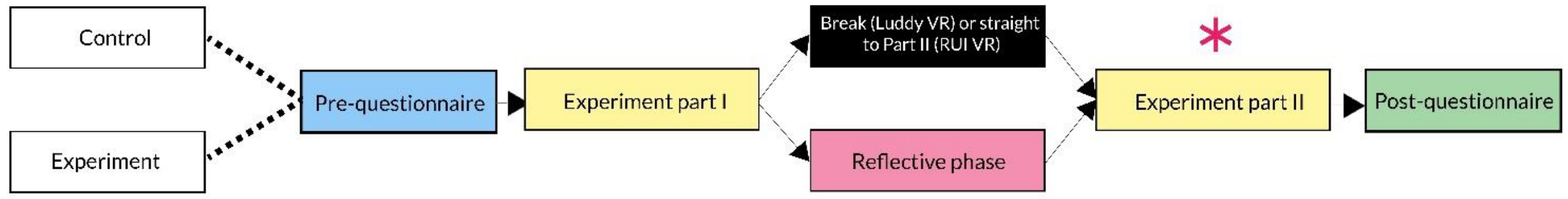

**Figure 1. General design of our studies with two cohorts. The purple asterisk marks the part of the experiment where we determined differences in post-treatment performance between the two cohorts.**

The studies featured two cohorts (control and experiment), both of which completed a pre-questionnaire about demographics and prior experience with VR and 3D, followed by a first set of tasks. The control group then either took a break (Luddy VR) or continued on to the next set of tasks. The experiment cohort, meanwhile, inspected their own data in a **Reflective phase** via a data visualization. Both groups then continued on to the second set of tasks and finally completed a post-questionnaire about their satisfaction and general impression of the application presented to them.

Table 1 compares the studies in this paper with regards to key design features. Both studies featured **two cohorts** (control and experiment). Likewise, both studies featured similar **visualization types** (map and line graph for RUI VR, map and bar graph for Luddy VR), similar **graphic symbols** (3D volume, line, and text), **graphic variables** (position, color hue, color saturation), and **interaction techniques** (filter, navigate). While we conducted both studies to provide data on the effectiveness of data visualizations for improving user performance in VR environments, there are differences in their design. The RUI VR study utilizes three **setups** (as it doubled as a comparison study between these three setups) while the Luddy VR study uses one. Similarly, the **graphic variable** of size was used for Luddy VR only, and the animate/replay and link and brush **interaction types** were applied only in RUI VR and Luddy VR, respectively. Finally, the **scale** of the reference system for Reflective phase visualization of each study was different (1:1 for RUI VR, 1:30 for Luddy VR).

Extensive documentation in the form of videos, images, and code can be found in the Supplementary Information and on GitHub (https://github.com/cns-iu/optimizing-performance-in-VR-using-DVL-FW). Since both studies involved benign behavioral interventions not posing greater than minimal risk, written consent by the participants was not required. The Institutional Review Board (IRB) of Indiana University approved our studies in an expedited review process. Participants provided their consent by agreeing to continue their participation after reading the study information sheet (SIS) at the beginning of the experiment. Additionally, we shared the SIS during the solicitation process via email, social media, and on monitors across the building of our school.

**Table 1. Side-by-side comparison of RUI and Luddy VR user studies.**

| | **RUI VR Reflective** | **Luddy VR** |
|---|---|---|
| **Setups (VR/Desktop)** | 3 (2D Desktop, VR Tabletop, VR Standup) | 1 (VR) |
| **Cohorts** | 2 (control, experiment) | |
| **Visualization types** | Map (VR setups), graph (2D Desktop) | Map, graph |
| **Scale of reference system** | 1:1 | 1:30 |
| **Graphic symbols** | Volume (VR setups), line (2D Desktop), linguistic/text (2D Desktop) | Volume, line, linguistic/text |
| **Graphic variables** | Position (3D, 2D), color hue, color saturation | Position, color hue, size, velocity |
| **Interactions** | Filter, navigate, animate/replay | Filter, navigate, link and brush |

### 2.1 RUI VR Study

We designed this experiment as a follow-up study to Bueckle et al. (2021), where 42 subjects across three setups (2D Desktop, VR Tabletop, and VR Standup), performed 14 increasingly difficult and then 30 identical cube-matching tasks either using a VR head-mounted device (HMD) while standing or sitting, or with a traditional 2D screen on a laptop. While the goal of that study was to compare accuracy, completing time, and satisfaction for these implementations, the goal of the study in this paper was to test whether data visualizations can be used to improve time, accuracy, and satisfaction in these VR vs. 2D Desktop implementations of the Registration User Interface (RUI). The RUI was developed to allow stakeholders in the Human Biomolecular Atlas Program or HuBMAP (Snyder et al., 2019) to register tissue blocks—i.e., to record the size, position, and orientation of human tissue into reference organs. Tissue mapping centers across the HuBMAP consortium have employed the RUI to register a total of 197 kidney, spleen, colon, lymph node, heart, lung, and thymus tissue blocks as of November 11, 2021 (Cyberinfrastructure for Network Science Center, 2021), with planned support for ca. 50 organs in the near future with the goal of constructing a Common Coordinate Framework (CCF) and a Human Reference Atlas for the human body at single-cell resolution. The basis of this study was formed by 44 cube-matching tasks, where users had to match the position and rotation of a white cube (tissue block) with a purple cube (target block) inside a 3D model of a human kidney. Study and task design are explained in detail in Supplementary Material, section 2 (Supplementary Text for RUI VR) as well as the narrated videos in the Supporting Information of a related publication (Bueckle et al., 2021) at https://github.com/cns-iu/rui-tissue-registration#video-demos-of-the-three-setups.

### 2.1.1 Study Design

The 84 participants in three cohorts were further divided into a control and an experiment cohort as shown in Figure 1,.

After filling out a pre-questionnaire about demographics and prior exposure to VR and 3D applications, subjects were assigned to one out of two cohorts (control, experiment) and one out of three setups. **2D Desktop** users completed their cube-matching tasks on a laptop with a 2D interface; **VR Tabletop** users used a VR HMD while seated at a desk (in physical and virtual space), and **VR Standup** users performed all tasks while standing in an area of ca. 3x3 meters (9x9 feet). After a brief tutorial, each user performed 14 increasingly difficult tasks (**Ramp-Up phase**) where the cubes became increasingly smaller, started farther apart, and with a greater angular difference. After this, subjects in the experiment cohort went through the Reflective phase, see below. Afterwards, each user completed a series of 30 identical tasks in the **Plateau phase** before finishing with a post-questionnaire.

In order to allow the users in the experiment cohort to inspect their own data from the Ramp-Up phase, we created a separate Unity application with the same base map, i.e., kidney and buzzer, as the Ramp-Up phase (for users in the VR Tabletop and VR Standup setups). 2D Desktop users were presented with a line graph visualization of the distance and angular difference between the tissue and target blocks. In this section, we outline what implementations of the Reflective phase looked like for each setup, the visual encoding, the interactivity, and the mid-questionnaire that concluded the Reflective phase before subjects continued with the Plateau phase. The Reflective phase consisted of two parts: an **intro** and a **main** part. In the intro, the user was shown a visualization of the best-performing subject in the control cohort of their setup in terms of completion times and position as well as rotation accuracy. A ~6 minutes tutorial (~3 minutes for 2D Desktop) introduced the goal of the Reflective phase, the visual encoding, the interactivity (for VR subjects), and prompted the user to derive strategies for faster and more accurate placement going forward. Subsequently, the user was shown their own data in the main part of the Reflective phase. While we measured the time spent in the Reflective phase by the user, we did not impose a minimum or maximum time limit on the user.

### 2.1.2 Research Questions and Hypotheses

**RQ1a**: Do users in the **experiment** cohort have a better performance in the **Plateau phase**, measured in accuracy and completion time, compared to the **control** cohort?
**RQ1b**: Do users in the **experiment** cohort have a higher satisfaction in the **Plateau phase**, measured in accuracy and completion time, compared to the **control** cohort?
**H1**: There will be a **significant** difference in completion time, accuracy, and satisfaction for Ramp-up and Plateau phases between control (without Reflective phase) and experiment group (with Reflective phase). However, this will only occur for the VR Standup and VR Tabletop setups, not the 2D Desktop users.

**RQ2**: In the Reflective phase, are metrics on user actions and interactive tool usage, measured through telemetry, correlated with higher performance in the Plateau phase?
**H2a**: More head rotations have a negative effect on completion times in the Plateau phase.
**H2b**: More head rotations have a negative effect on distance (higher position accuracy) in the Plateau phase. This may be due to high-performing users feeling more comfortable in 3D environments in general, and VR specifically, enabling them to move around their own data more fluently in the first place.

### 2.1.3 Reflective Phase

In this section, we outline the Reflective phase implementation for all three setups.

#### 2.1.3.1 2D Desktop

2D Desktop users were shown a line graph, see Supplementary Figure 6. On the x-axis, we plotted the elapsed time in seconds as well as task numbers; on the y-axis, we added two scales: distance between the two blocks (left side, measured in Unity scene units) and the angular difference (right side). Additionally, we inserted vertical dot-dash lines to indicate the end of one task and the beginning of the next one. This static visualization was created using R and Shiny after loading a CSV file with data from the subject's Ramp-Up phase, created in Unity previously at runtime.

#### 2.1.3.2 VR Tabletop

In the VR setups, we used the inherently spatial reference system of the virtual environment to produce 3D dot density maps, encoding the HMD, hand, and tissue block positions over time. Figure 2 shows the Reflective phase setup for a user in the VR Tabletop setup. They were seated at the same virtual and physical tables as they were during the Ramp-Up phase, ensuring a 1:1 mapping. They were allowed to move around during the Reflective phase and inspect their data form multiple angles.

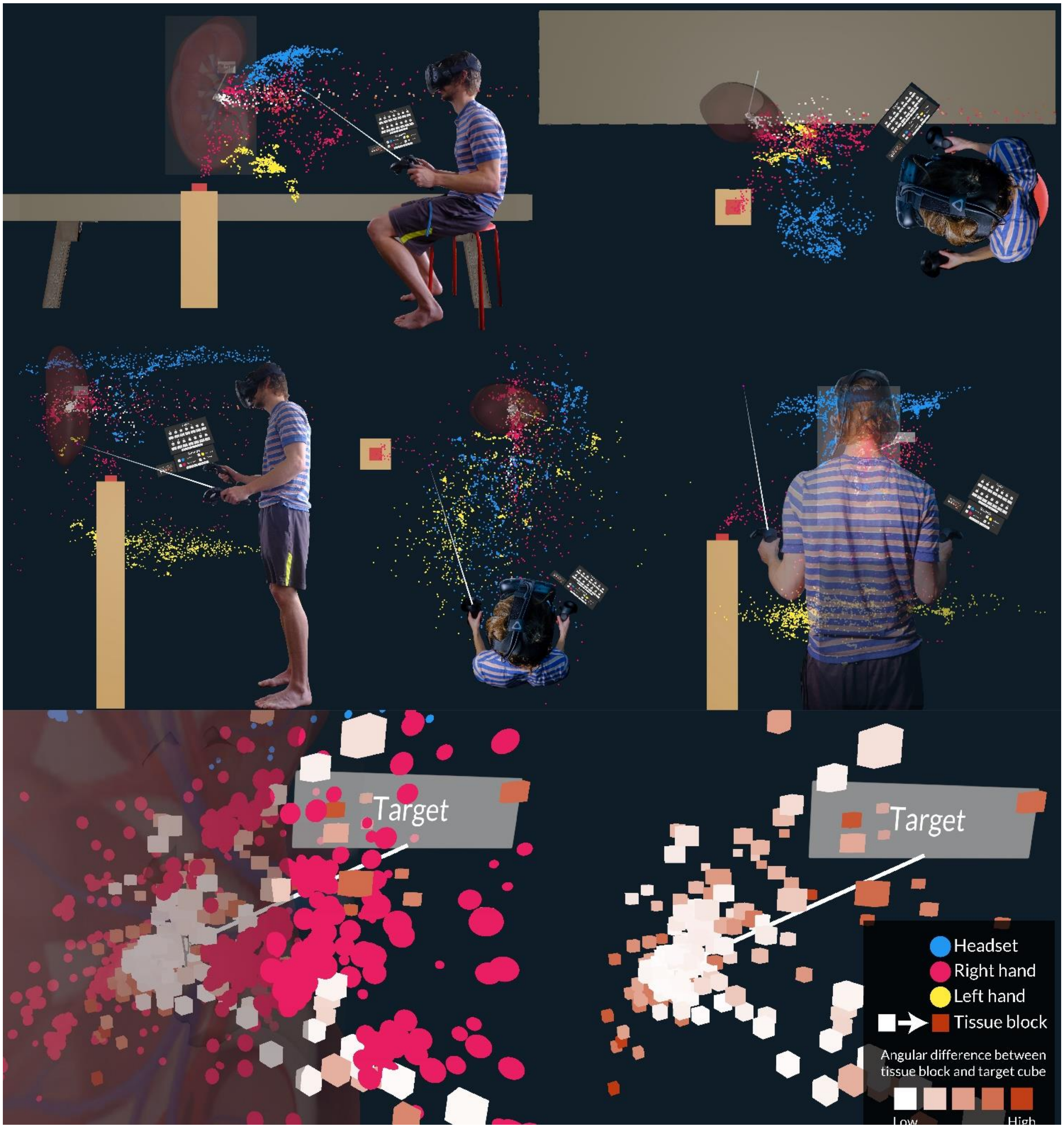

**Figure 2. Reflective phase setup for VR Tabletop and VR Standup. (A) side view (VR Tabletop). (B) top view (VR Tabletop). (C) Side view (VR Standup). (D) Top view (VR Standup). (E) Back view (VR Standup). (F) Close view of tissue block positions over time near the target block with all data visible. (G) Same view without kidney and data for right hand.**

### 2.1.3.3 VR Standup

Similar to subjects in the VR Tabletop setup, VR Standup users explored first the data of the best-performing users from the control cohort and then their own. Figure 2 contains side (C), top (D), and back view (E). Just like in VR Tabletop, users in VR Standup were allowed to explore the 3D dot density map freely by walking around the space while using the kidney and buzzer as a base map.

### 2.1.4 Visual Encoding

Like the Reflective phase implementations for the three setups, the visual encodings applied to each setup differed as well. In this section, we describe the graphic symbol and graphic variable pairings (Börner, 2015) used to encode the telemetry and task performance data from the Ramp-Up phase.

#### 2.1.4.1 2D Desktop

In the 2D Desktop setup, the graphic symbol line encoded two data records: position accuracy, expressed as the distance between the tissue block and the target block over time, and the angular difference, i.e., the difference in rotation between the two blocks, expressed as a single value from 0 (same rotation) to 180 (diametrically opposed rotation). The graphic variables x-y position and color hue encode position and rotation accuracy, respectively. Additionally, various linguistic and pictorial symbols provide additional information to the user: axes are properly labeled; a note at the bottom of the graph indicates the height of the kidney so that reading the position accuracy values becomes easier; vertical dot-dash lines mark the beginning of a task and the start of the next; white gridlines help with reading values off the y-axes. As a temporal visualization, the x-axis contains the elapsed time in minutes and seconds (since the end of the tutorial task).

#### 2.1.4.2 VR Tabletop and VR Standup

For the two VR setups, we used a straightforward visual encoding scheme for the user's HMD and controllers: blue for the HMD, pink for the right controller, yellow for the left controller, white to orange for the tissue block over time. The graphic symbol volume, together with the graphic variable color saturation, encodes the angular difference between the tissue block and the target block and was indicated to the user in a legend (see Figure 2G). For all graphic symbols, the graphic variable x, y, z-position encoded the x, y, z-position if the corresponding device (HMD, controllers) or virtual object at a given moment in time. The resulting visual encoding allowed users to quickly identify areas of concentrated activity. Frequently visited areas of space were thus indicated by a higher density of dots (like the area around the target block, see Figure 2F).

### 2.1.5 Interactivity

The aforementioned areas of concentrated activity were visualized in an aggregate view that the user encountered when they first entered the stage. By default, all data was shown; as a consequence, patterns in the movements were easier to spot, but the large number of dots generated by the user over the course of the Ramp-Up phase also led to visual clutter (see Figure 2F). To allow the user to remove various layers of data through filtering of their choice, we implemented two interactions: **filter** and **animate**.

The area around the target block position shown in Figure 2F and Figure 2G tended to amass a large amount of data records due to frequent user activity when fine tissue block placement was performed. A mix of pink dots and white-orange cubes visualized the user's right hand placing the tissue block while minimizing the angular difference. By using their controller, the user could remove parts of the base map (the kidney) as well as parts of the data overlay by a series of features: **graphic symbol type**, **time stamp**, and **task number**. In Figure 2G, for instance, the user has removed the graphic symbol for the right hand (pink dots) as well as the kidney to declutter the display around the target.

Figure 3B shows a screenshot of the interactive legend presented to the user on top of their right controller. It consisted of three sections: **graphic symbols**, **tasks**, and a **static legend** for the angular difference. The graphic symbols part allowed the user to turn parts of the data overlay on and off by entire types of data records encoded by these symbols. Checkboxes enabled the user to **filter** by graphic symbol type and task number. Lastly, we allowed the user to show and hide data records by time stamp. Specifically, we implemented a time slider on top of the user's left controller (see Figure 3A). The time slider consisted of a slider area with a play head, similar to what one would find in a video editing program. The user could move the play head along the slider by putting their thumb onto the trackpad on the left controller.

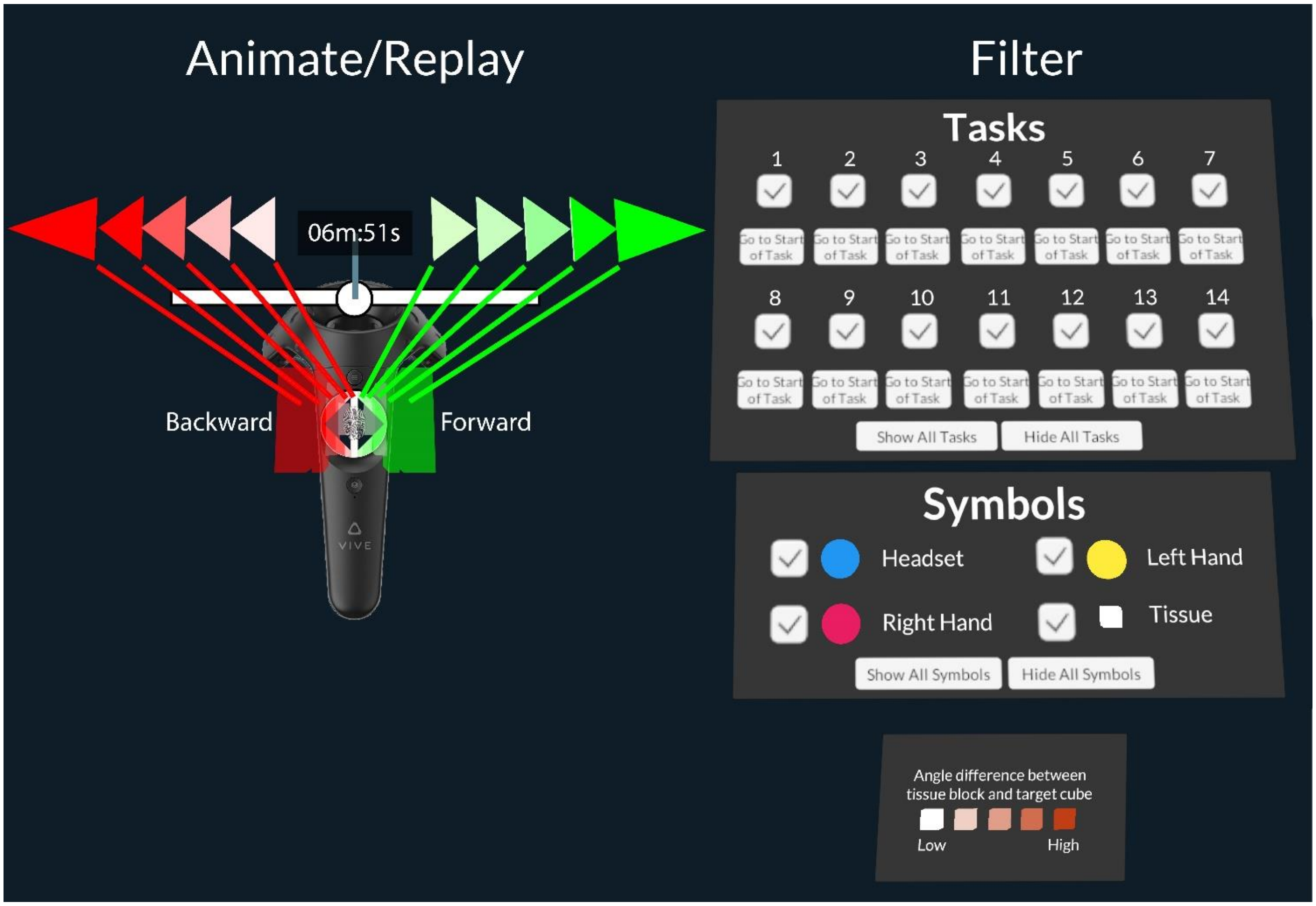

**Figure 3. A: Filter menu to turn parts of the data overlay on and off by task number or graphic symbol type. B: Close-up of the left controller with play head speed zones.**

Placing the thumb onto the right and left half of the trackpad let the user skip forward and backwards through the dataset by time stamp, respectively. This allowed the user to replay the dataset at various speeds depending on their horizontal distance from the center of the touchpad. Additionally, they could activate a fast-forward and fast-backward mode when additionally pressing the trigger button on the back of the controller. To indicate to the user the current speed at which they were skipping through the dataset, green or red arrows were displayed next to the current time stamp (in minutes and seconds since the beginning of task 1). In parallel, 3D green and red blocks were displayed over the touchpad in the user's virtual view.

The combination of task, graphic symbol, and time stamp filter formed an interactive system that let the user quickly isolate individual tasks, replay tasks, as well as isolate specific elements of the scene (such as the tissue block), and enabled them to switch between aggregate and focused views in a way that felt natural and immersive. When implementing this interactive system, we used the native C# event system to broadcast any changes to the UI elements to the graphic symbols in the scene. The graphic symbols had behaviors attached to them that then evaluated whether all conditions were met for them to be shown or hidden.

### 2.1.6 Metrics

To measure user performance, we defined three metrics: completion time, position accuracy, and rotation accuracy. Users could finish tasks by pressing a 2D button (2D Desktop) and by hitting a virtual, red buzzer button (VR Tabletop, VR Standup). Completion time was captured in seconds per task, captured from the frame during which the button was pressed to the next time it was pressed. We defined position accuracy as the distance between the two cubes (centroid distance), and rotation accuracy as the angular difference in degrees, both assessed at the time of task submission. We captured satisfaction in the post-questionnaire via a 5-point Likert scale, ranging from -2 (not at all satisfied) to 0 (neutral) to +2 (very much satisfied). A detailed overview of how we assessed and compared position accuracy and rotation accuracy between the three setups can be found in a related publication (Bueckle et al., 2021) as well as the Supplementary Material, section 2.4 (Metrics).

## 2.2 Luddy VR study

Following the RUI VR study, we designed a second experiment, involving the navigation of virtual buildings. Our goal was to test whether the completion time for traversing virtual buildings can be improved using data visualizations in VR. To conduct the study, we used a 3D model of Luddy Hall, the home of the School of Informatics, Computing, and Engineering at Indiana University in Bloomington, IN, USA. The model was designed by Philip Beesley Architect Inc. (http://www.philipbeesleyarchitect.com/). Since this model was built as a scaffold for a public art piece to be installed by the architect's studio (https://cns.iu.edu/amatria.html), it came in two parts: a simpler version of the entire building where most structures were just hinted at, without any materials; and a more detailed version of just the atrium of Luddy Hall. For the entire experiment, the user spent time only in the atrium of Luddy Hall. Detailed screenshots are shown in Figure 4.

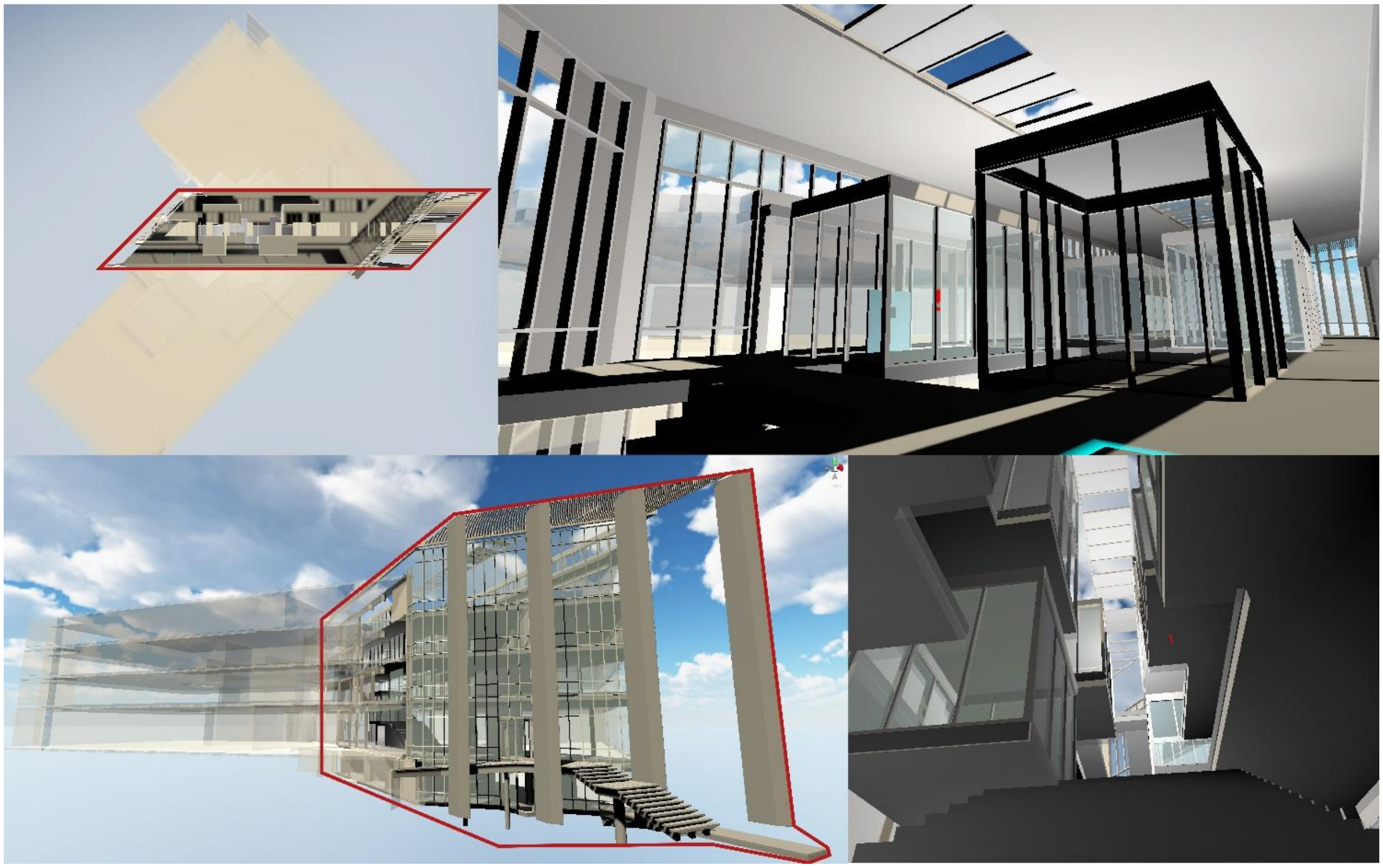

**Figure 4. Exterior and interior shots of the Luddy Hall model. A: top view with atrium highlighted orange. B: 4th floor near star case, start point for all navigation tasks. C: side view, atrium highlighted. D: view from first floor up the central stair case.**

Like for RUI VR (see section 2.1), the goal of this study was to identify whether we could observe performance improvements between a control and experiment cohort. Both cohorts performed a series of navigation tasks using a VR HMD and controllers for two sets of 24 tasks (including tutorials) with the option of three different navigation methods, with a break in-between (control) and a Reflective phase to inspect their performance from the first trial in order to formulate strategies for improvement in trial 2. We implemented three common navigation choices in VR (walking, teleporting, free-flying), which we explain in more detail in section 2.2.4.

### 2.2.1 Study Design

When arriving at the research site in Luddy Hall, the subject was asked to sit down at a table with a laptop running a survey. The survey began with a study information sheet before presenting a pre-questionnaire to obtain information about the subject's demographic background as well as prior experience with VR, video games, 3D applications in general, data visualizations, and their familiarity with Luddy Hall. Following that, each subject put on the VR gear. During the following VR Trial 1, they performed a total of 24 tasks in four rounds.

Following that, the control group took a break from the study. The research facilitator encouraged them to stand up and walk around the research area. The experiment group, on the other hand, stayed in VR and was presented with a Reflective phase, where they saw their own data visualized as 3D trajectories across a miniature version of the building. We describe this in more detail in section 2.2.5.

After the break (control) or the Reflective phase (experiment), subjects from both cohorts sat down at the laptop to fill out a mid-questionnaire, where we asked them how many tasks they had completed in total, how many floors the building had, and more questions. The mid-questionnaire is discussed in section 2.2.6. Then, all subjects donned the VR gear again for VR Trial 2, where they repeated the same 24 tasks from VR Trial 1, but without any audio tutorials. We excluded the tutorial tasks from the data analysis.

Finally, all subjects completed a post-questionnaire, where we asked them to rate their own performance, state their preference for the navigation tasks, and indicate their satisfaction with their performance. After successful completion of all parts of the study, each subject was remunerated with a $20 gift card.

### 2.2.2 Research Questions and Hypotheses

For our study, we aimed to answer the following research questions and provided the following hypotheses.

**RQ1**: Is there a difference in completion time between the control and experiment cohorts during trial 2?
**H1**: The experiment cohort achieves significantly lower completion times than the control cohort during VR Trial 2.

**RQ2**: Is there a difference in the rate of change in completion time from trial 1 to trial 2 between the two cohorts? That is, when computing the differences in completion time per trial and per subject, and then compare these values between the cohorts, is there a significant difference?
**H2**: The experiment cohort achieves significantly larger changes in completion times between trial 1 and trial 2.

**RQ3**: When asked questions about the tasks and the virtual building after taking a break (control) and completing their Reflective phase (experiment), is there a difference in score between the two cohorts?
**H3**: The experiment cohort achieves higher scores in the mid-questionnaire than the control cohort.

**RQ4**: What are the preferred choices of navigation methods during the last round of tasks?
**H4a**: Subjects prefer teleporting when finalizing a task within sight of the start position.

**H4b**: Subjects prefer free-flying when finalizing a task out of sight of the start position.
**H4c**: Subjects prefer walking just as they finalize a task.

**RQ5**: Is there a difference in self-reported satisfaction between the two cohorts at the end of the experiment?
**H5**: There will be no significant difference in satisfaction between the cohorts.

### 2.2.3 Task Difficulty

At the core of this study were two sets of 24 navigation tasks in VR (see Figure 5B). During each of the first three rounds, only one navigation method was possible, starting with walking, then going to teleporting, and ending with free-flying. In the fourth round, the subject could choose which navigation method they wanted to use, and they could change it at any time. The first task in every round served as a mini-tutorial where a pre-recorded voice explained the scope and goal of the experiment as well as the controls of the currently active navigation method to them.

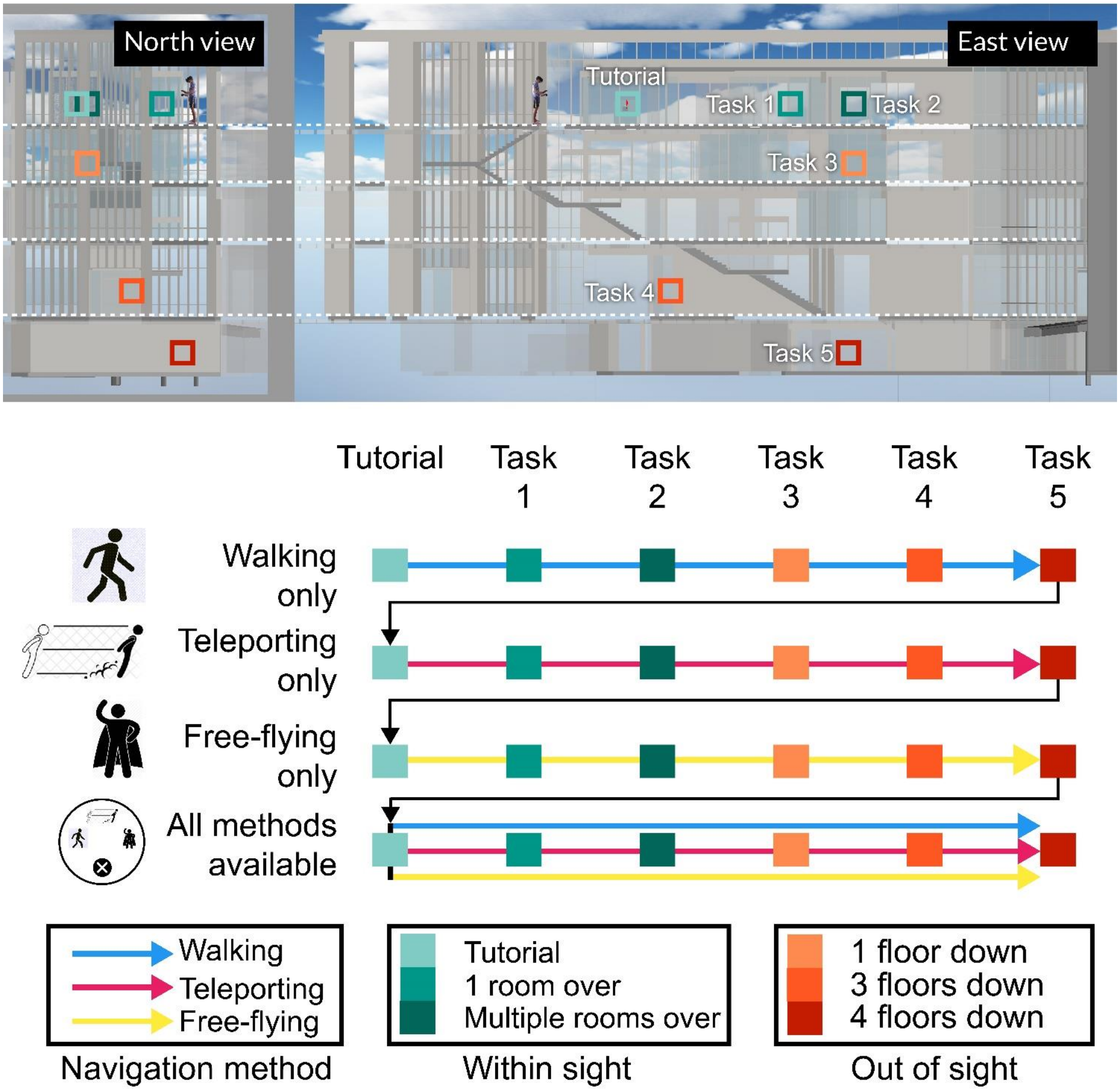

**Figure 5. This distribution of tasks across Luddy Hall and task sequence. A: the five navigation tasks (plus tutorial) at their locations in two aligned cross-section views of the building. B: all 24 tasks in sequence with color-coded difficulty level and possible navigation methods.**

All 24 tasks entailed navigating from a fixed start position at the top of the fourth floor (see the human figure in Figure 5A). The target positions for tasks 1 and 2 were on the same floor and within sight, in two study rooms. Tasks 3-5 were on increasingly lower levels, with task 3 being inside another study room, task 4 being in a small office space under the staircase on the ground floor, and task 5 being in a classroom in the lower levels of the building. We used the increasing distance between the start position and a task target position as well as whether a target was within sight or out of sight to increase the difficulty over time. The distances between the start position and tasks 1-5 were approx. 11.6, 17.4, 24.3, 18.7, and 27.1 meters in a straight line, with distances increasing consistently between tasks. Task 4 was an exception (with 18.7 meters), but it was four floors down (on the ground floor) and out of sight of the start position, making up for the slightly smaller distance than the preceding tasks 3.

Inside each task room, there was a flat, blue panel with a white sphere inside it, with a diameter of about 10 cm (see Supplementary Figure 16C and D). The center of the sphere was about 120 cm above the floor. Below the sphere, a text panel instructed the user to finish the task by holding their controller against the sphere for one second. When the user arrived at a task, they had to locate the panel, approach it, interact with the sphere; subsequently, they were transported back to the start position.

To indicate to the user where to navigate next, a red exclamation mark was placed as a waypoint inside each task room. We used a custom shader to ensure that the waypoint was rendered on top of any other surfaces in the scene (see Supplementary Figure 16B). To help the user locate the task room once in sight, we added a flurry of around 4000 purple, twinkling particles to each room.

### 2.2.4 Navigation Methods

We present the calculations for speed and direction for each navigation method in detail in Figure 6.

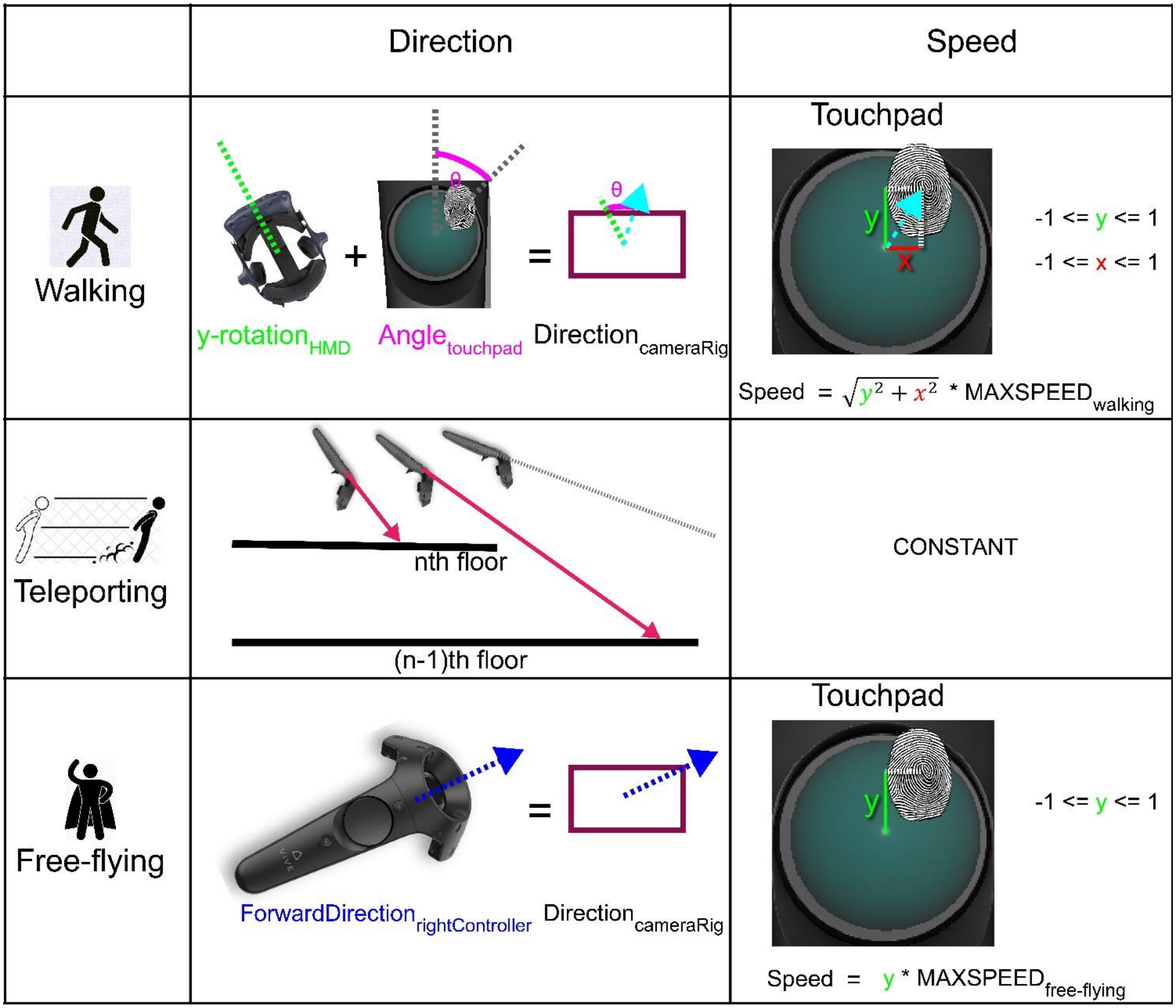

**Figure 6. Advanced illustration of how direction and speed are calculated from user input, for each navigation method.**

**Walking** was the navigation method most closely imitating real-world locomotion on foot. The user was bound by gravity; when leaping over an edge, they could fall until they hit a surface with a collider. It was possible to walk up and down the stairs traversing the center of the atrium. We added invisible walls to the outside of the atrium to prevent users from falling to infinity. These walls were only active during sections when the only possible navigation method was walking.

The user controlled the speed and direction of their walking with their right controller and their HMD. The thumbpad on the controller is represented as a unit circle in the SteamVR SDK, and touchpad positions are given as (x, y) positions (where -1 <= x <= 1 and -1 <= y <= 1). The walking speed was then determined by the product of the distance of the touch position from the center of the touchpad and the maximum speed possible in the walking mode (2.5 meters per second). The walking direction, on the other hand, was determined by two angles: the y-rotation value of the HMD and the angular distance between the position of user's thumb and the y-axis on the touchpad (see Figure 6). This setup allowed the user to walk independently of the direction of their gaze if needed by simple use of their thumb. Modulating the speed with the thumb ensured that faster and slower velocities were possible, and by moving their thumb into the lower quadrants of the thumbpad, the user could also walk backwards if needed.

**Teleporting** allowed the user to traverse distances within sight with a click of the touchpad on their right controller. In order to teleport, the user had to point their controller towards a suitable surface, and a purple ray would then indicate their target position were they to execute the teleport.

To facilitate teleporting, we embedded a ray caster alongside the z-axis (forward vector) of the 3D representation of the controller, which persistently sent rays into the scene. When arriving at a teleport destination, the user's virtual camera rig was automatically adjusted such that the bottom plane of the user's camera rig was on the ground. The ray always returned a hit location on the first suitable surface it encountered; teleporting thus only allowed the user to navigate within sight.

Finally, **free-flying** allowed the user to travel through space with increased freedom. Neither gravity nor physical barriers were in their way. Using their right controller, the user could manipulate speed and direction. Speed was defined as the product of the y-value of the user's touch position on the thumbpad ($-1 \le y \le 1$) and the maximum speed for free-flying, which was set to 3 meters per second. The flying direction corresponded to the forward vector of the right controller, i.e., wherever the user pointed their right hand. Flying backwards was also possible.

When **all** navigation methods first became available, the navigation method active previously was deactivated, and we did not give the user a standard method to prevent any biases towards a specific navigation method. The user could switch between all three at any point through putting a radial menu over their left controller, triggered by touch. The currently active navigation method was displayed as text on a small panel over the left controller.

### 2.2.5 Reflective Phase

Users in the experiment cohort got to inspect their own data in a mix of 3D and 2D data visualization inside VR (see Figure 7). To that end, we created a 3D trajectory visualization, consisting of a dot density map of user positions over time, and a bar graph on a 2D panel over the user's left controller. This panel also contained a set of checkboxes to turn the data for individual tasks on and off. This setup presents a mix of spatial and abstract data visualization in one comprehensive VR interface for testing if users can utilize these types of data visualizations to improve their performance in VR Trial 2.

The visualization was created at runtime using a custom C# script, reading in data from a CSV file generated while the user completed VR Trial 1. The script iterated through every row in the dataset, instantiated the appropriate graphic symbol depending on the navigation method chosen, and added a data component for each graphic symbol that could later be used for the interactive legend to turn parts of the data overlay and off depending on user input.

To familiarize themselves with the visualization and the controls, we presented the Reflective phase to the user in two parts: First, we showed them data from a high-performing user in the control cohort; simultaneously, they were listening to a 4-minute audio tutorial introducing them to the base map, the data overlay, and the controls while outlining the goal of the Reflective phase: to identify strategies in their data to improve their own performance. Then, after the tutorial was done and the user decided they had had enough practice, we loaded their own data in the visualization.

**Base map and data overlay**: The miniature model of Luddy Hall was around 75 cm tall, floating in front of the user such that the roof of the building was around 140 cm above the floor. The user could free-fly around the model so that they could inspect the model from all angles, independently of their physical height and range of motion.  The entire visualization was resized at a scale of 1:30.

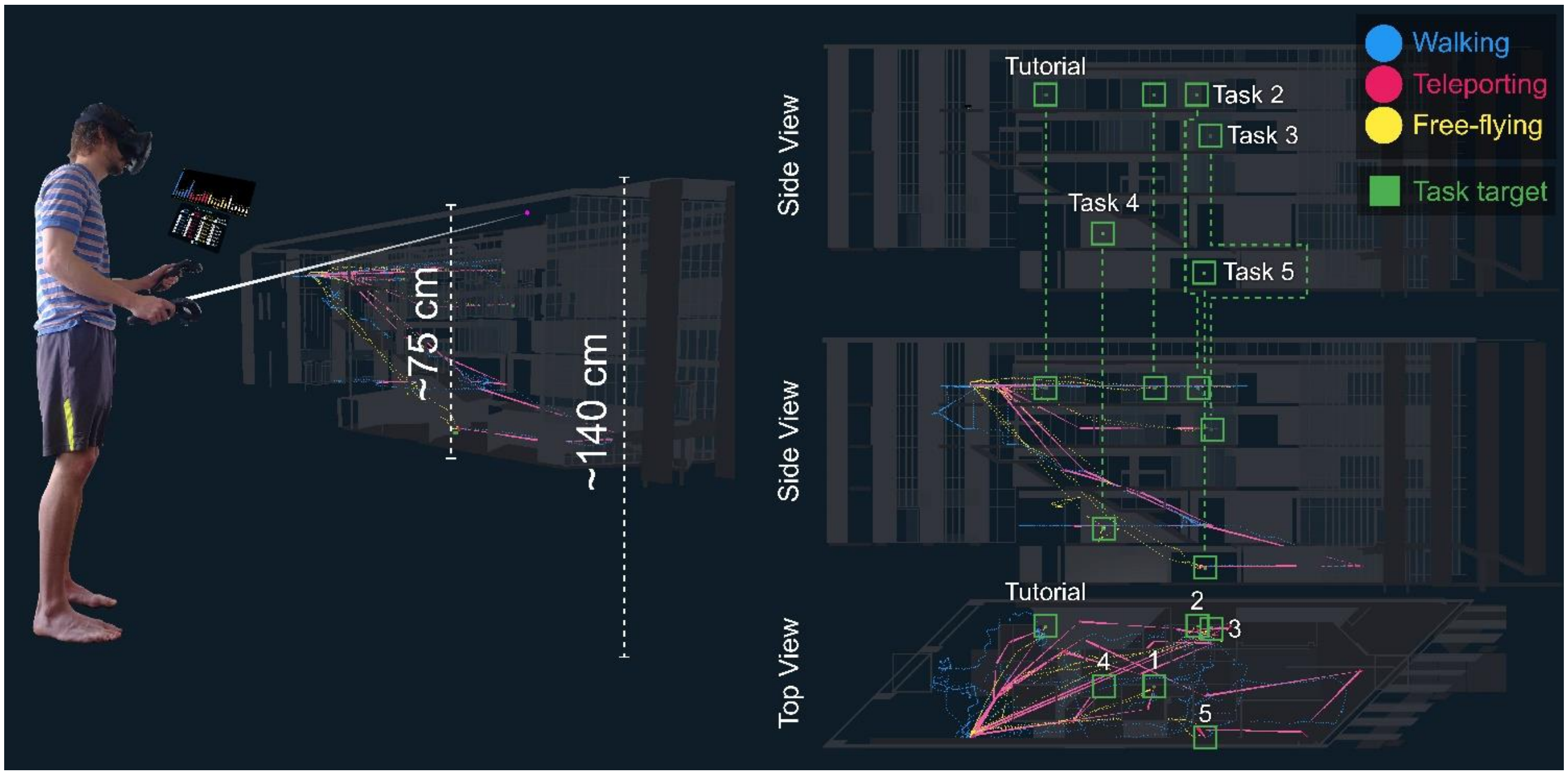

**Figure 7. A: A user with model and data visualization during Reflective phase. B: Three views of a user's data in the Reflective phase. Green cubes in the visualization show the destination for each task. Blue, pink, and yellow visualize walking, teleporting, and free-flying, respectively.**

**Visual encoding**: We chose color hue to represent navigation method, and x, y, z-position of the dot to encode the user’s position. For teleporting, we encoded each teleport as a pink line, starting with the same width as the dot, and then thinning out evenly towards the teleport target.

**Bar graph for completion times**: In order to give the user quick insights into their performance, we displayed a 2D bar graph on top of their left controller with completion times for all 24 tasks (see Figure 8).  On the y-axis, we showed the completion time in seconds; the color of the bar encoded the navigation method possible during the task; and the x-axis contained the task number. On top of each bar, the completion time in seconds (rounded to one decimal) was displayed.

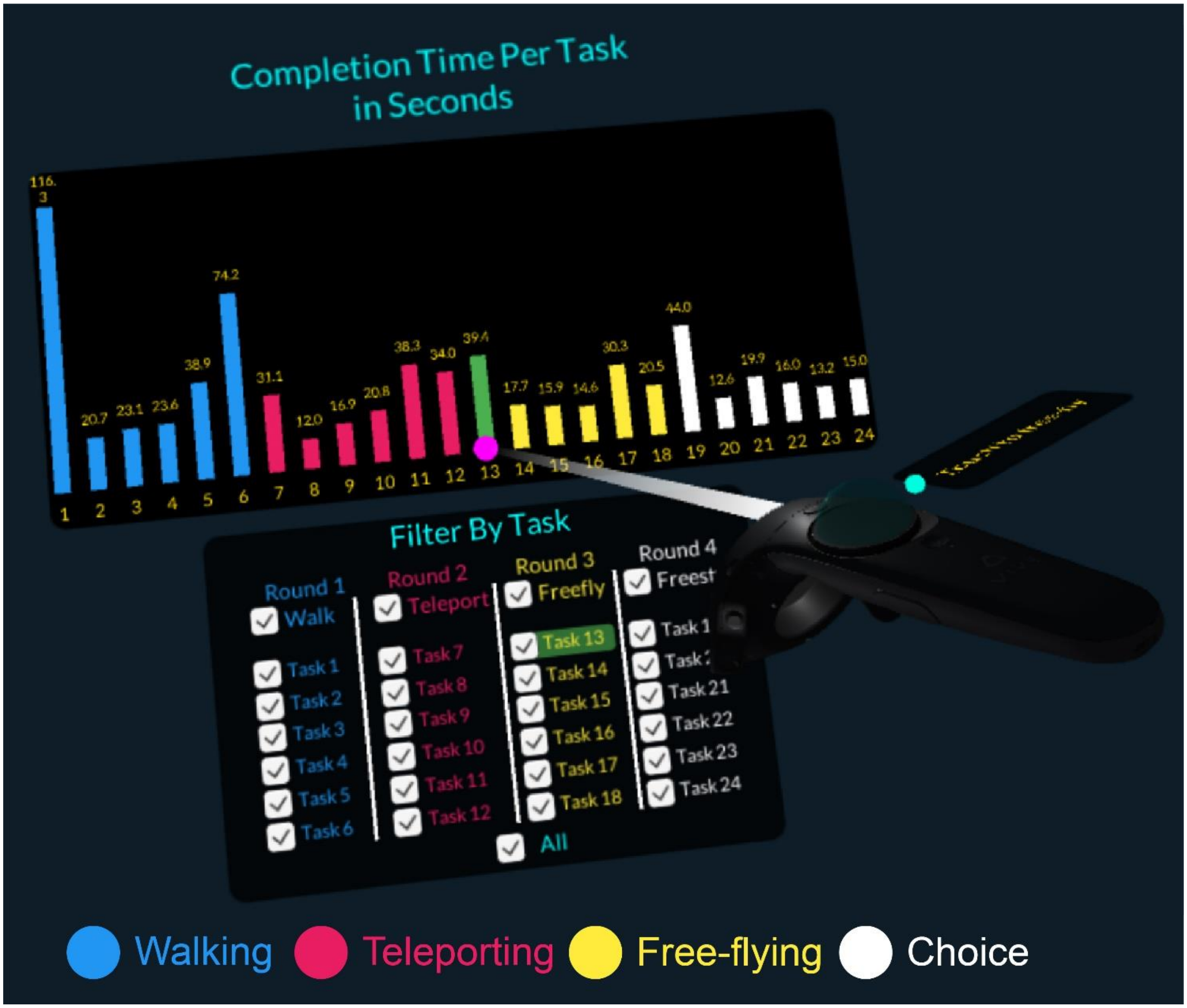

**Figure 8. The interactive legend and bar graph visualization for the completion times of all 24 tasks in VR Trial 1. Note the link and brush functionality as the user is hovering over the bar for task #13.**

**Interactive legend**: The bar graph functioned as part of an interactive legend that helped the user turn parts of the data overlay on and off. To allow the user to focus on the tasks they wanted to explore, we implemented a link and brush functionality. The subject could use their right controller as a pointer (see Figure 8). When hovering over a bar, both the bar and the corresponding checkbox to turn the corresponding graphic symbols were highlighted green. The user could then identify particularly long or short completion times and then inspect the corresponding trajectories in the Luddy Hall model. A checkbox to show or hide all data was located at the bottom of the interactive legend.

### 2.2.6 Mid-questionnaire

Before trial 2, all subjects answered 10 questions about the features of the building (e.g., number of floors), the tasks they performed (e.g., how many tasks total). The goal of the mid-questionnaire was to test whether there was a difference in spatial memory and understanding between the two cohorts.

### 2.2.7 Metrics

The metric to assess performance was **task completion time** in seconds, measured from the frame when the user gained control of their movement to the frame when they had touched the virtual submit button for one second (see Supplementary Figure 16D), at which point the timer was reset. Another metric for our data analysis was the user's **mid-questionnaire score**. Finally, we included a self-reported **satisfaction** score at the end of the post-questionnaire, where the user had to indicate whether they felt satisfied after the experiment, using a 5-point Likert scale, ranging from -2 (not at all satisfied) to 0 (neutral) to +2 (very much satisfied).

## 2.3 Apparatus

We ran both experiments on an Alienware 17 R4 with a 17.3" display, running Windows 10 with an Nvidia GTX 1070 and 16 GB RAM. We used an HTC Vive VR HMD with controllers on SteamVR (Valve Corporation, 2021) in a play area of around 9x9 feet (3x3 meters). We used the Camtasia (TechSmith, 2020) screen-recording software as well as a Logitech C930e webcam to capture the user's action with audio and video, both in VR and the physical world.

## 2.4 Statistical Approach

In this section, we describe the data analysis for our two studies.

### 2.4.1 For RUI VR

First, we describe how we analyzed the influence of tool usage during the Reflective phase on performance in Plateau phase tasks and satisfaction. As **dependent** variable, we chose the difference (improvement) between mean performance variables in the 14 increasingly complex Ramp-Up phase tasks and the Plateau phase tasks such that

$$\Delta performance_var_i = performance_var_plateau_i - performance_var_ramp_up_i$$

The mean performance variables for the Ramp-Up tasks for each subject $i$ have a significant Pearson correlation with their Δ counterparts — for completion time and angular difference (rotation accuracy) only, not for centroid distance (position accuracy):

**Table 2. Pearson correlations between variables in RUI VR. Note that centroid distance is the measure for position accuracy (distance between centroids of the tissue block and target block) and angular difference is the measure for rotation accuracy (difference in orientation between the tissue block and target block).**

| Variable 1 | Variable 2 | Pearson correlation | p-value |
|---|---|---|---|
| $\text{mean_task_completion_time}_{\text{ramp_up}}$ | Δtask_completion_time | - 0.840 | $p < 0.001$ |
| $\text{mean_centroid _distance}_{\text{ramp_up}}$ | Δcentroid_distance | 0.020 | $p = 0.920$ |
| $\text{mean_angular_difference}_{\text{ramp_up}}$ | Δangular_difference | - 0.493 | $p = 0.008$ |

Because there are only 28 observations (14 for VR Tabletop, 14 for VR Standup), it is not possible to test the effect of all Reflective phase variables on the delta-performance measures at once. Therefore, for each performance measure, a model is estimated:

$$\Delta_{\text{performance}} = \propto\ + \beta_1\text{reflective_variable}\ +\ \beta_2\text{ramp_up_performance}\ +\ \varepsilon$$

This way, we one can control for Ramp-Up performance while estimating the effect of changes in the single reflective phase variables. **Ordinary Least Squares** (OLS) estimation was used with clustered standard errors on the level of experiment conditions. Estimations with the dependent variable of self-reported satisfaction were estimated without control variable. The regression model was estimated using the “lm” (DataCamp, 2021b) command from the R “stats” package (R Core Team, 2021), which fits a model using OLS. The clustered standard errors are computed with the “coeftest” command (DataCamp, 2021a) from the R package “lmtest” (Hothorn et al., 2021). We used an OLS estimation with clustered standard errors on the level of the VR condition because we cannot assume that the error terms of our estimations are uncorrelated to the subjects' conditions. We report our results in section 3.1.

### 2.4.2 For Luddy VR

For the Luddy VR study, we conducted a series of **Welch’s Two-Sample t-tests** between the control and experiment cohorts to compare completion times for VR Trial 1 and 2, satisfaction, motion sickness, and mid-questionnaire score. Additionally, we computed **Pearson's product-moment correlations for** motion sickness with mid-questionnaire score and satisfaction. We report these results in section 3.2.

## 3 Results

This section contains the results of our user studies. We evaluate whether there was a measurable difference between the control and experiment cohorts before discussing potential reasons for our findings.

### 3.1 RUI VR Study

Supplemental Figure 23 shows the demographic make-up of the participants in a series of stacked bar graphs. Notably, out of 84 subjects, there were 45 subjects who identified as male, 38 as female, and one subject who preferred not to answer. In terms of age, 52 participants were between 21 and 30 years old. 81 subjects were right-handed, 3 were left-handed. No subjects experienced color blindness. 44 subjects had prior experience in 3D modeling, with no differences between the cohorts. The majority of subjects (50) had not played first-person shooters in the past 12 months.

We performed a Kruskal-Wallis-Test (McKight and Najab, 2010) for distribution of gender and task completion time, position accuracy, rotation accuracy, and satisfaction, and found **no significant differences**. Likewise, we conducted a Mann-Whitney-U-Test for distribution of handedness (left vs. right) task completion time, position accuracy, rotation accuracy, and satisfaction without detecting significant differences, either (see Supplementary Table 3).

### 3.1.1 Influence of Reflective Phase on Performance and Satisfaction

Figure 9 shows a series of boxplots for position accuracy, rotation accuracy, completion time, and satisfaction across both cohorts in all three setups.

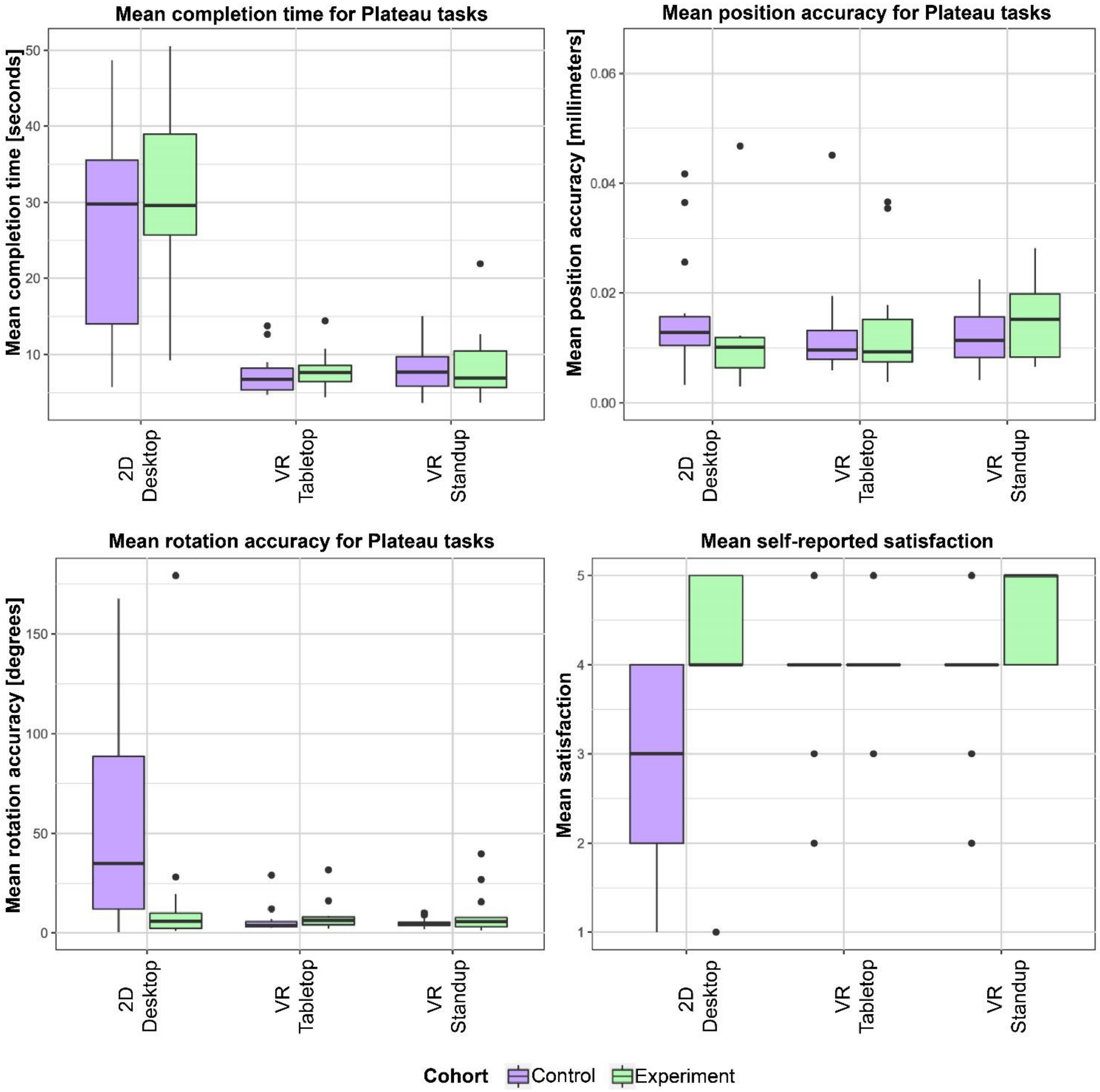

**Figure 9. Mean completion time (A), position accuracy (B), rotation accuracy (C), and satisfaction (D) for Plateau tasks in all setups for both cohorts.**

We found **no difference** between the cohorts for completion time, position, and rotation accuracy for the two VR setups. However, in the experiment group of the **2D Desktop** setup, we found a **significantly** higher rotation accuracy (median experiment group: 5.806 degrees, IQR: [2.216 - 9.949], median control group: 34.98 degrees, IQR: [12.109 - 88.457], Mann-Whitney-U-Test, **p = 0.031**) as well as a **not significant** difference in position accuracy (median experiment group: 0.01, IQR: [0.006 - 0.012], median control group: 0.013, IQR: [0.01 - 0.016], Mann-Whitney-U-Test, **p = 0.085**). This means that the Reflective phase (line graph) for the 2D Desktop users did indeed help users outperform the Desktop users in the control cohort. However, we found no difference in terms of completion time for 2D Desktop. Based on these findings, we have to **reject H1**.

Surprised by these findings, we performed further analyses for the users' self-reported **satisfaction**, and found a **significantly higher** feeling of satisfaction in the experiment group of the **2D Desktop** setup (median experiment group: 4, IQR: [4 - 5], median control group: 3, IQR: [2 - 4], Mann-Whitney-U-Test, **p = 0.004**) and in **VR Standup** (median experiment group: 5, IQR: [4 - 5], median control group: 4, IQR: [4 - 4], Mann-Whitney-U-Test, **p = 0.006**). In section 3.1.2, we examine the user behavior in the Reflective phase of the VR Standup setup to determine what factors may have contributed to this higher level of satisfaction with one's performance.

### 3.1.2 Metrics During Reflective Phase and Influence on Performance in Plateau Phase

We further wanted to understand the relationship between metrics for user behavior as well as interactive tool usage and performance in the Plateau phase and self-reported satisfaction. In this section, we focus on whether there are **VR behavior traits** in the Reflective phase that have an effect on the performance in the Plateau phase (for VR users only). This could help us pinpoint what specific elements of the Reflective phase could be adjusted to improve user performance in real-world VR training. We thus discuss **effects** between metrics during the Reflective and Plateau phases for the experiment cohort to understand how behavior in the Reflective phase influences performance in the Plateau phase in order to answer **RQ2** (see **)**. An explanation of our data analysis using Ordinary Least Squares (OLS) can be found in section 2.4.1.

**Table 3. Regression table for effects of tool usage and behavior in Reflective phase on performance in Plateau phase. + desirable, - not desirable, ***significant at the 1% level, **significant at the 5% level, *significant at the 10% level**

| **Variable 1 (Reflective)** | **Variable 2 (Plateau)** | **Effect size** | **Result** |
|---|---|---|---|
| Total time spent (Reflective phase) | Completion time (Plateau phase) | **-0.00285*** ** | + |
| Total time spent (Reflective phase) | Satisfaction | **-0.00003*** ** | - |
| Time without kidney visible (Reflective phase) | Centroid distance/position accuracy (Plateau phase) | **0.00876** ** | - |
| Time without kidney visible (Reflective phase) | Angular difference//rotation accuracy (Plateau phase) | **3.53305* ** | - |
| Time without kidney visible (Reflective phase) | Satisfaction | **-0.63312*** ** | - |
| Total head rotations around y-axis (Reflective phase) | Angular difference//rotation accuracy (Plateau phase) | **-0.00010* ** | + |
| Total head rotations around y-axis (Reflective phase) | Satisfaction | **0.00001*** ** | + |

First, we find that spending **more time** in the Reflective phase has a **significant negative effect** on task **completion time** in the Plateau phase, without jeopardizing **position** or **rotation accuracy**. However, it has a **negative effect** on the **satisfaction**. This might be, because the Reflective phase presents an analytical mode as opposed to the almost playful rest of the experiment where users actually get to interact with virtual objects. Additionally, many users, when confronted with their own data for an extended period of time, may have discovered their performance in the Ramp-Up phase to be lacking.

Second, not seeing the **base map**, i.e., the kidney, for extended periods of time in the Reflective phase has a **significant negative** effect on **satisfaction** and a **positive effect** on both centroid distance and angular difference in the Plateau phase, resulting in **lower position** and **rotation accuracy.** This is a wholly undesirable. Seeing one's data without the proper context seems to be a major issue not only for performance but also enjoyment of the entire VR experience. The integrity of the base map or reference system thus seems conducive to accuracy and user satisfaction.

Third, we identified metrics that enhanced both **accuracy** and **satisfaction** metrics. The **total number of degrees of head rotation around the y-axis** had a negative effect on angular difference (**rotation accuracy)** and a positive one on **satisfaction, without jeopardizing completion time** and centroid distance **(position accuracy)**, prompting us to **reject both H2a and H2b**. We thus conclude that encouraging users to look around in their environment is conducive to gaining visual insights.

## 3.2 Luddy VR Study

For the Luddy VR study, we recruited 71 subjects via email lists, social media, and word-of-mouth. While running the experiment, two subjects had to abort their participation during VR Trial 1 due to motion sickness. One more subject had to stop during VR Trial 2 for the same reason. This left us with a total of 68 subjects for data analysis (34 per cohort). Subjects spent an average of **43.5** (**control, SE = 1.24**) and **60.4 minutes** (**experiment, SE = 2.08**) in the study from entering the research area to leaving it. For all analyses, we omitted the tutorial task, i.e., the first task in every round where users were introduced to the navigation method for that round in the first trial and the corresponding tasks in the second trial.

### 3.2.1 Demographics

30 of our subjects identified as female and 38 as male. The majority of subjects (40) were between 21 and 30 years old; further, 20 were between 18 and 20, six were between 31 and 40, and two were between 51 and 60 years old. The overwhelming majority were English native speakers (54). 65 subjects were right-handed; 3 were left-handed. Around half of the subjects (32) stated that they had no vision impairments; 26 were near-sighted. Participants were allowed to wear glasses under the HMD or contact lenses as needed. 67 subjects indicated that they were not color-blind; one subject preferred not to answer that question.

In terms of prior experience with VR, video games, and 3D applications in general, the overwhelming majority of subjects had used VR before (51); out of these, 37 had used it rarely, nine occasionally, and five often. The HTC Vive, Vive Pro, or Cosmos was the most used VR system, indicated by 21 subjects. Over two thirds of subjects (43) said that they played video games in the past 12 months, mostly on smartphones or other handheld devices (27). Also, 27 subjects had played first-person shooters. Further, 36 subjects had used 3D software before, such as Rhino (6), AutoCAD (5), and Unity (5). When asked whether they would say that they were familiar with Luddy Hall, 10 subjects strongly agreed, 22 somewhat agreed, 10 neither agreed nor disagreed, nine somewhat agreed, and 17 strongly disagreed. We also asked our subjects to state their familiarity with six basic visualization types (tables, charts, graphs, maps, trees, and networks) and did **not find any correlation** between the self-reported familiarity scores with six visualization types and the correct answers in the mid-questionnaire.

### 3.2.2 Performance Improvement

To answer RQ1 (whether there was a difference between the control and experiment cohorts for completion times in VR Trial 2), we isolated each subject's task completion times (minus the four tutorial tasks), leaving us with 20 observations by subject (680 by cohort). To ensure that observations where users needed excessive amounts of time did not obstruct the validity of our analysis, we removed 39 observations from the control cohort (**m = 50.7 seconds, SE = 1.95 seconds**) and 38 observations from the experiment cohort (**m = 47.13 seconds, SE = 1.59 seconds**). We then performed a Welch's Two-Sample t-test with the remaining observations. This showed that there was **a significant difference** in task completion times between the two cohorts during VR Trial 2 (**$m_{control}$= 16.44 seconds, $SE_{control}$ = 0.30, $m_{experiment}$ = 15.44 seconds, $SE_{experiment}$ = 0.27, $t$ = 2.465, $p$ = 0.014**). We were then interested in determining whether this difference was caused by the Reflective phase or whether subjects in the experiment cohort were naturally more able in VR. We thus compared the completion times for VR Trial 1 after removing 42 and 40 observations from the control and experiment cohorts, respectively. A Welch's Two-Sample t-test then showed **no significant difference** between the two cohorts for completion times in VR Trial 1 (**$m_{control}$ = 20.75 seconds, $SE_{control}$ = 0.41, $m_{experiment}$ = 21.26 seconds, $SE_{experiment}$ = 0.45, $t$ = -0.843, $p$ = 0.399**). We thus **reject the null hypothesis for H1** (the experiment cohort achieves **significantly lower** completion times than the control cohort during VR Trial 2). Figure 10 contains a collection of boxplots for the completion times per task and navigation method, separated by cohort.

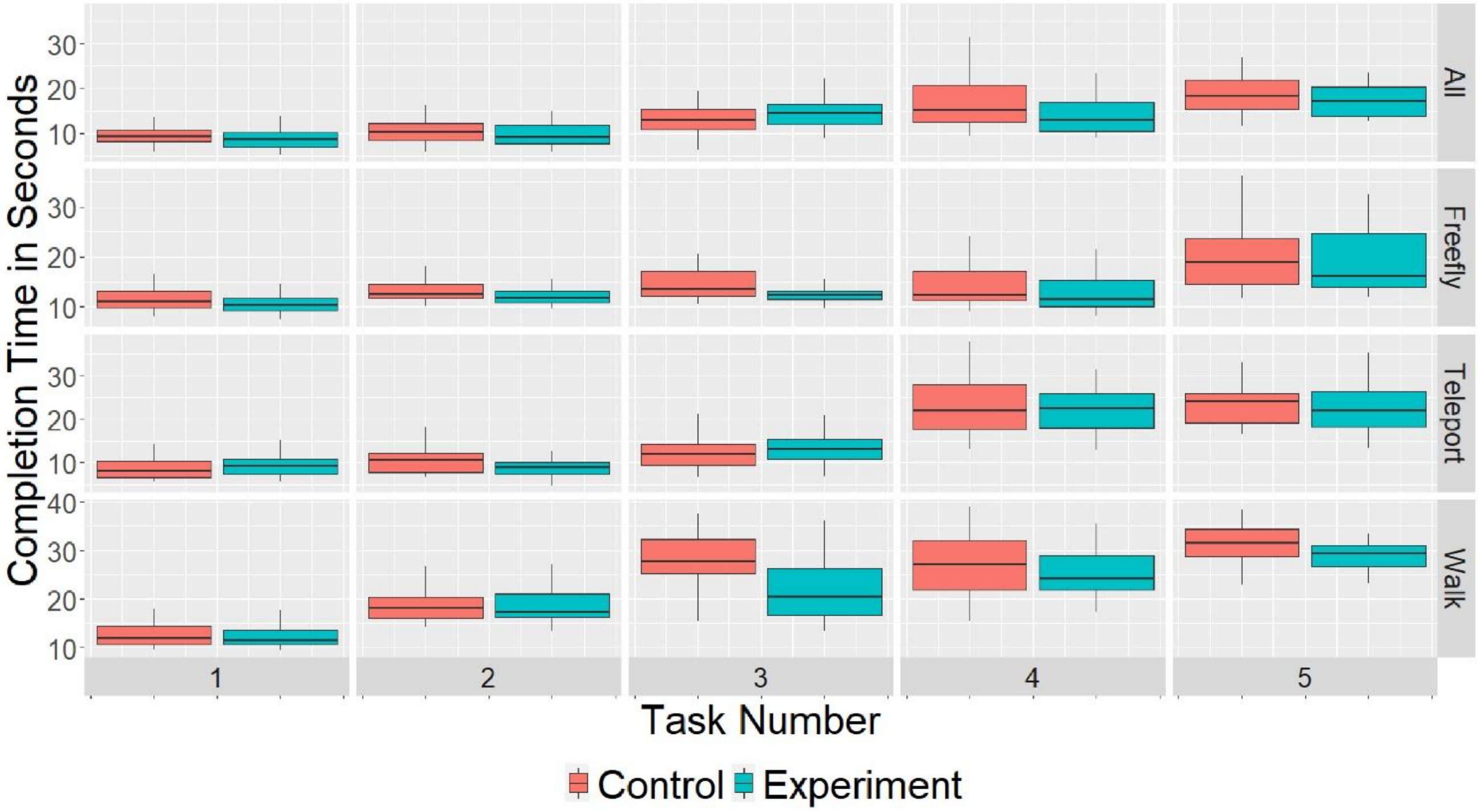

**Figure 10. Faceted boxplots of completion times by task number (horizontal) per round (vertical).**

Further, while it was to be expected that both cohorts would improve their times during VR Trial 2 to some degree (due to the learning effect), we wanted to identify the difference in the rate of change for completion times. We found a **significant difference** between the two cohorts (**$m_{control}$= -5.38 seconds, $SE_{control}$ = 0.446, $m_{experiment}$ = -7.48 seconds, $SE_{experiment}$ = 0.498, $t$ = 3.146, $p$ = 0.002**).

This means that users in the experiment cohort improved by more than 2 seconds compared to users in the control cohort (on average). We thus **reject the null hypothesis for H2**.

### 3.2.3 Mid-questionnaire Score

In addition to checking whether our intervention helped the experiment cohort achieve lower completion times than the control cohort in VR Trial 2, we compared the scores from the mid-questionnaire, where we asked our users to answer questions about the navigation tasks they performed with regards to the spatial layout of Luddy Hall. In a t-test (**$t = -2.703$, $p = 0.009$**), we found that experiment users (**$m = 5.71$, $SE = 0.371$**) performed **significantly better** than control users ($m$ = **4.29, $SE = 0.367$)**. We thus **reject the null hypothesis for H3**. Figure 11 shows violin plots for the total task scores for both cohorts along with jittered points for all 68 scores (and both medians).

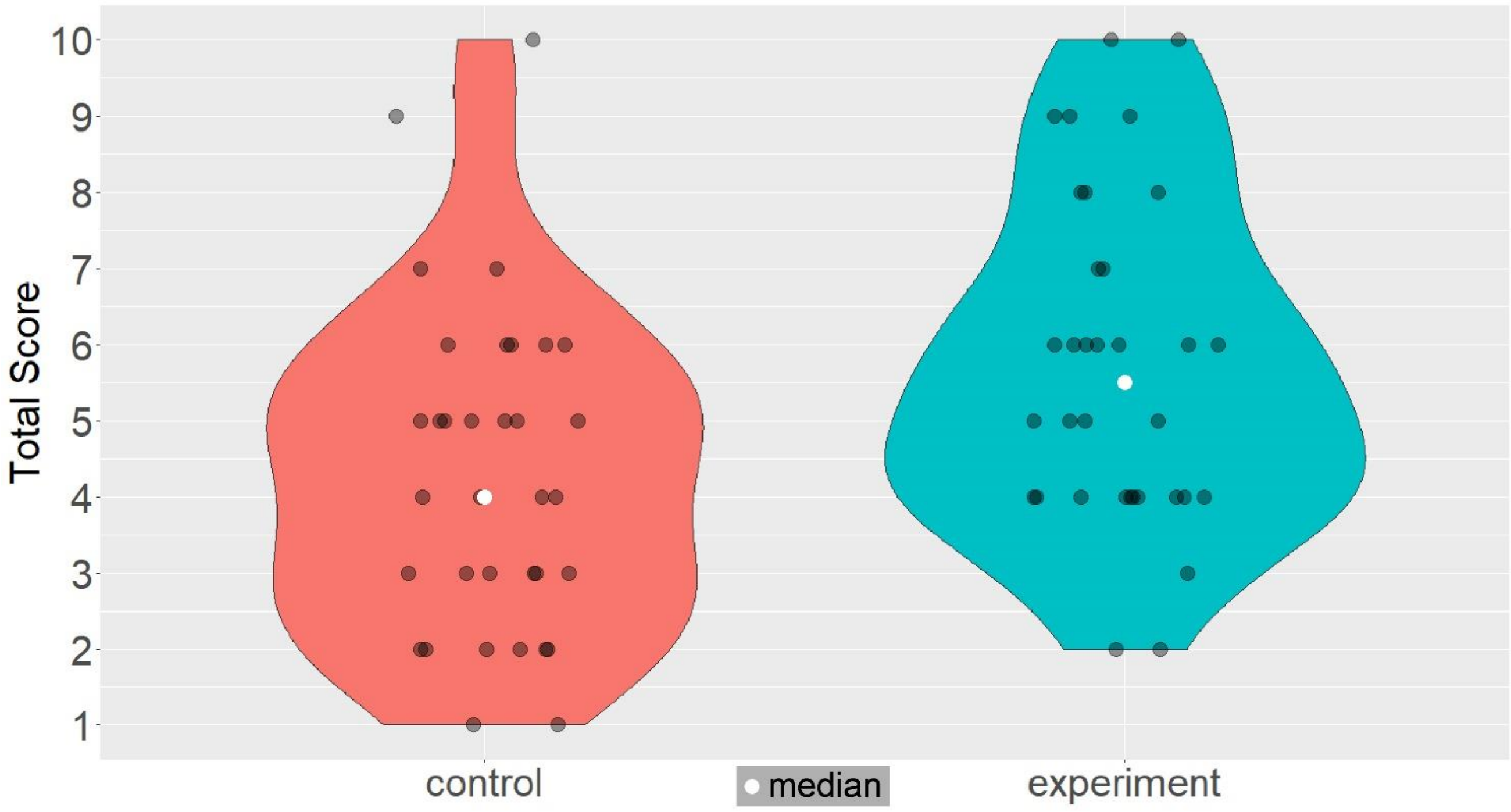

**Figure 11. These violin plots show a significant difference in the mid-questionnaire score between the two cohorts.**

### 3.2.4 Choice of Navigation Methods

During the last round of tasks in VR Trials 1 and 2, the subjects could switch between navigation methods at will. In RQ4, we wanted to check what navigation methods subjects would employ when completing these tasks. We had hypothesized that subjects would use teleport to reach targets within sight of the start position (**H4a**), free-fly for targets out of sight of the start position (**H4b**), and walking at the very end as a means to get to floor level (**H4c**).

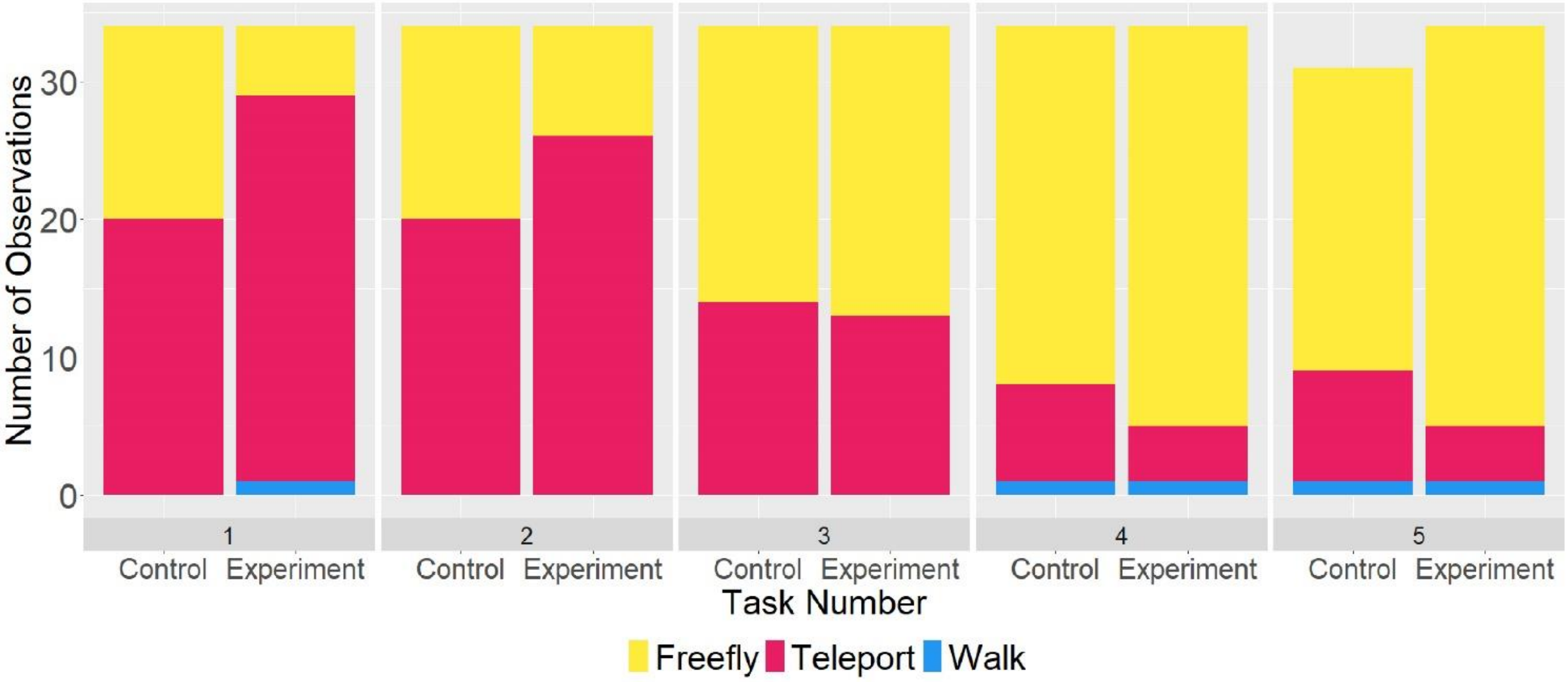

**Figure 12. Navigation methods selected by subjects at task submission during VR Trial 2. Three outliers for the control cohort were removed for Task 5.**

Figure 12 shows a bar graph of the last logged navigation method for the 5 tasks in the last round of VR Trial 2 for all 68 subjects (minus three outliers in the control cohort), yielding 337 observations. The tendency for all subjects to end tasks 1 and 2 with teleporting (both are on the same floor as the start position) becomes apparent; for tasks 3 through 5, users preferred free-flying as a way to traverse larger distances quicker (and go through walls thanks to the lifted physical restrictions when using free-flying). This "winning strategy" is more pronounced within the experiment cohort than the control cohort. While we can **reject the null hypothesis for H4a** (teleporting for targets within sight) and **H4b** (free-flying for targets out of sight), only a minority of users ended tasks by walking, prompting us to **reject H4c** (walking is preferred for ending tasks quickly).

### 3.2.5 Satisfaction

In terms of satisfaction, we found **no significant difference** between the cohorts, and thus **reject the null hypothesis for H5**. Both had a high mean satisfaction on a 5-point Likert scale (**$m_{both}$ = 4.29, $SE_{control}$ = 0.107, $SE_{experiment}$ = 0.143, t = 0, p = 1)**. Additionally, we found **no significant Pearson's product-moment correlation** between a user's score in the mid-questionnaire and their reported satisfaction (**r = 0.217, p = 0.08**). 40 subjects reported at least a little motion sickness (**m = 1.63, SE = 0.069**) on a scale from 1 (no motion sickness) to 3 (very much motion-sick). However, it appeared to have **no significant correlation** with the total score (**Pearson's product-moment correlation, r = 0.081, p = 0.51**). Similarly, we found **no significant difference** in motion sickness between the two cohorts **($m_{control}$ = 1.59, $SE_{control}$ = 0.086, $m_{experiment}$ = 1.67 $SE_{experiment}$ = 0.109, *t* = -0.635, *p* = 0.528**).

### 3.2.6 Post-questionnaire Results

In our post-questionnaire, we aimed to identify which navigation method was the most popular. To that end, we asked subjects to rank all three navigation methods from most to least favorite. The absolute majority of subjects (**n = 35**) preferred **free-fly** to teleport and walk; just over a quarter (**n = 19**) favored teleport over free-fly and walk, and only a minority of subjects liked walk the most (**n = 5**). Note that no single subject preferred walk over free-fly and teleport (in that order). Further, we found that all users, regardless of cohort, liked the VR experience as indicated by high means on a 5-point Likert scale for different aspects: overall (**m = 4.57, SE = 0.07**), hardware (**m = 4.49, SE = 0.09**), and instructions (**m = 4.59, SE = 0.07**).

### 3.2.7 Reflective Phase Feedback

Similarly, we aimed to measure how subjects in the experiment cohort perceived the interactive tools at their disposal during the Reflective phase using a 5-point Likert scale, and found that subjects overwhelmingly found them useful: filters and checkboxes (**m = 4.32, SE = 0.14**), color coding (**m = 4.68, SE = 0.13**), the bar graph of completion times (**m = 4.29, SE = 0.17**), and the ability to fly around the miniature base map of Luddy Hall (**m = 4.26, SE = 0.16**).

The subjects then indicated how efficiently they thought they had navigated the 3D space while only one navigation method was possible (again using a 5-point Likert scale). These numbers reflect the rankings of the navigation methods discussed earlier: Users overall found that they did not perform efficiently with regards to walking (**m = 2.29, SE = 0.18**) while teleporting (**m = 3.74, SE = 0.18**) and free-fly (**m = 4.12, SE = 0.20**) were ranked higher.

# 4 Discussion

In this paper, we presented two user studies with Reflective phases involving different interactions and base maps. We designed the Reflective phases in our studies by consulting the DVL-FW (Börner et al., 2019), specifically its typologies for interactions, graphic symbols, and graphic variables. Notably, for the Luddy VR study, users that experienced the Reflective phase had **significantly better** performance in terms of task completion time and better scores in the mid-questionnaire. We saw no such improvement for users in the VR setups of the RUI VR study, only for the 2D Desktop users who were presented with line graphs. In this section, we examine the differences between the Reflective phases in these two studies as far as the DVL-FW typology is concerned.

## 4.1 Comparison of Reflective Phase Implementations

The key differences between the Reflective phases are listed in Table 1 in section 2. Based on these, we conclude that a variety of factors accounted for our results.

**Visualization type and scale of reference system**: First, the visualization in the RUI VR Reflective phase presented the user's data on a 1:1 scale. The building in the Luddy VR study, however, with its geospatial layout, waypoints, stairs, rooms, and floors, was at a scale of 1:30. We conclude that this difference between the two reference systems provided more value to the Luddy VR users for devising strategies to improve their performance. There is currently no **scale** typology in the DVL-FW that could capture this difference in a meaningful way.

Second, Luddy VR users were given an **additional visualization type** in the form of a (bar) graph with their completion times per task, allowing them to quickly identify patterns, trends, and local maxima and minima in their performance via a more abstract data visualization. In RUI VR, on the other hand, the completion time was not explicitly shown, and could only be inferred from the time stamp on the time slider (for VR users). This level of abstraction for visually non-explicit information probably yielded a disadvantage.

**Graphic symbols and graphic variables**: Both Reflective phase implementations had a set of graphic symbols and graphic variables in common. Mainly, both used volumes (spheres, cubes) to encode the user's position (and the position of their hands) as well as the location of the tissue block over time (RUI VR) and the task destinations (Luddy VR). For the implementation of the more traditional 2D visualizations (line graph and bar graph), lines were used in both studies (though just for 2D Desktop in the RUI VR experiment). Both implementations also make use of linguistic symbols, specifically text, to denote task numbers (both), completion times (2D Desktop in RUI VR, all of Luddy VR), and parts of the interactive legends. In terms of graphic variables, there were differences. Both Reflective phases used position (3D for VR visualizations and 2D for line graph and bar graph) and color hue to encode the input device visualized (RUI VR) or the navigation method chosen (Luddy VR). In RUI VR, however, we also used color saturation to indicate the angular difference between the tissue and target blocks (and thus the rotation accuracy), while in Luddy VR, we used the graphic variable of size to encode the user's performance metric (completion time) in the bar graph. Finally, in Luddy VR, we employed velocity in two ways: when visualizing the vector between a user's teleporting start and end point, and to show head and hand movement direction when replaying a dataset in RUI VR. All graphic symbol and graphic variable encodings were received and understood well based on mid- and post-questionnaire data. Generally, we conclude that the 3D trajectories that made up the Reflective phase visualizations illustrated the performance for virtual navigation tasks better, because trajectories through virtual environments are a type of data overlay familiar to most people through navigation systems in cars, on phones, and in video games. The head and hand movements in the RUI VR study may have been too abstract for users to derive viable strategies.

**Interactions:** Over the two studies, we implemented four interaction types: **filter**, **navigate**, **animate**, and **link and brush**. **Filter** was implemented in both studies: The user could turn parts of the data overlay on and off via checkboxes in the interactive legend, using a 3D pointer. Likewise, users could **navigate** across the visualization by virtue of wearing VR equipment, allowing them to see the data from different angles. **Link and brush** and **animate/replay** were unique to Luddy VR and RUI VR, respectively. Luddy VR users could brush over a bar in their graph of completion times and then saw the corresponding menu entry highlighted that let them turn on and off the underlying data for that task. RUI VR users could play back their own data at different speeds by using the time slider, thus creating an animation of their own movement and tissue block manipulation over time. However, it appears that this interaction type was less useful than expected.

It may have been hard for users to identify and select a playback speed that yielded insights about their completion times to them, and offering predetermined playback speeds may have been a better design choice. Additionally, playback for a task worked better when other data was turned off, thus **avoiding clutter**, and the steps needed to hide all data, jump to the time stamp of a task, and then playing it back at an insightful speed, required a series of actions on the user's part that may have been challenging for many users. Also, we assume that the animate/replay interactivity did not help users identify completion times and potential optimizations properly, because completion time had to be derived from looking at time stamps, not directly via an auxiliary visualization like in Luddy VR. Link and brush for Luddy VR users, on the other hand, created a visual connection between the bar graph and the dot density map of their trajectories, thus allowing the user to evaluate their performance with this additional, derived data rather than having to infer their performance.

### 4.2 Design Implications

The **filter**, **link and brush**, and **navigate** interactions proved to be valuable for Luddy VR users, who were able to achieve faster completion times than their counterparts in the control cohort. While we did not find that the Reflective phase helped RUI VR experiment users (in the VR setups) performed better than control users in the Plateau phase, we did find significant effects by metrics in the Reflective phase on performance in the Plateau phase and satisfaction. These effects could help identify ways to nudge users to a better performance by tweaking parts of the Reflective phase. Subsequently, we describe which of these insights are actionable for the future design of interventions using the DVL-FW.

We identified one metric that had both a favorable effect on rotation accuracy and satisfaction: **head rotation around the y-axis**. Similarly, we found that many variables that describe the Reflective phase experience made the entire VR experiment less satisfying for subjects, among them total time spent, time slider usage, and time without the base map, which also had a detrimental effect on position and rotation accuracy (see ). It thus appears that a beneficial improvement would be to **bring the data to the user**, e.g., by scaling down the reference system of the visualization (like for Luddy VR), thus minimizing the need for covering a wide area with their entire body. A more refined Reflective phase would thus encourage the user to make use of their head as a "camera" to inspect the data from different angles in a 3D perspective while not moving their entire body around. At the same time, this could also shorten the time spent in this analytical mode.

Additionally, we determined that less aggregation leads to happier users. This means that another good tweak would be to **avoid aggregated data views** where possible, especially at the start of the Reflective phase where all the data was turned on by default. Of course, it is also possible that the visualization type, together with the aggregated data view, was the confusing and less satisfying element for many RUI VR users. If they had been presented with a 2D dot density map instead of the 3D + VR version of our experiment, an aggregated view could have revealed more patterns (like the 2D dot density maps in Supplementary Figure 13).

It would also be possible to design a Reflective phase for RUI VR as a **mix of VR visualizations and traditional, 2D visualizations** like in Luddy VR. The 3D dot density map is an advanced visualization type, and even if users extracted insights from their own data, it may not have empowered them to act on these insights. This would also relieve the user from the need to derive performance variables such as the completion time while trying to memorize more successful strategies.

Yet another, more far-reaching change would be to more closely entangle any reflection about one’s own data with the actual task hand rather than outsourcing it to a separate Reflective phase application. **Immediate visual feedback** could be given when the cube-matching task is done to encode accuracy, possible following some of the design goals of fluid interaction, such as “provide immediate visual feedback on interaction” (Elmqvist et al., 2011). Additionally, because the tasks were performed in VR, **haptic feedback** via the VR controllers could be used to indicate position and rotation accuracy during the tasks, shortening the user’s time in the Reflective phase.

## 4.3 Limitations

We acknowledge several limitations to these studies. First, while both RUI VR and Luddy VR contained a Reflective phase, the overall **study design** was slightly different. In Luddy VR, both cohorts completed the same set of tasks twice, with the intervention in between. In RUI VR, the Ramp-Up phase contained a different set of tasks than the Plateau phase, so it was not possible to compute within-subject improvements for these users.

Second, the **types of task** between the two studies were different (cube-matching vs. movement), which influenced what information users needed to retrieve from the Reflective phase. For example, RUI VR users had to identify ways to balance their efforts between position accuracy (with a focus on arm movements) and rotation accuracy (with a focus on hand movements), all without neglecting completion time. Luddy VR users, on the other hand, experienced a more mediated interaction in that they could move their entire body through the virtual space via simple button clicks and touches. As a result, the navigation methods demanded less physical movement and input. Likewise, data from the RUI VR cube-matching tasks was richer as users did not only have to take position into account when determining new strategies but also rotation of the tissue block. As a result, improving performance for RUI VR users was **more challenging**. This would have been hard to any user, regardless of expertise with visualizations and VR, be it due to lack of expertise, spatial ability, confidence in virtual environments, or Data Visualization Literacy. This is highlighted by the fact that users in the Desktop setup (with the simple, static line graph) were able to significantly improve their position and rotation accuracy while feeling more satisfied using a rather simple visualization.

Third, for the experiment cohort in the RUI VR study, participation in this study was a lot more **time-consuming** than for the control cohort, potentially yielding a benefit to the latter. From reading the study information sheet at the beginning of the pre-questionnaire to answering the final question of the post-questionnaire, RUI VR subjects spent an average of **3627.21 seconds** (**SD = 789.38 seconds) or ~60.45 minutes** on the entire experience versus an average of **1811.67 seconds** (**SD = 840.1 seconds**) **or ~30.19 minutes** for control subjects. This resulted in an average **difference of slightly over half an hour** between these two cohorts. This is in stark contrast to Luddy VR, where experiment users needed **3624.82 seconds** (**SD = 717.5 s**) **or ~60.41 minutes** versus **2608.41 seconds** (**SD = 428.46 s**) **or ~43.47 minutes** for the control cohort (on average). This resulted in an average difference of just **under 17 minutes**, which is only ~57% of the time difference between the cohorts in the RUI VR study. Also, subjects who went through the Reflective phase in the RUI VR study put on and then took off the HMD four different times, switching between HMD and the laptop with the instructions every time. In the future, the research design for similar interventions could be more streamlined.

Finally, while the telemetry data from the HMD and VR controllers allowed us to model a user's behavior for our data analysis, it did not allow us to draw conclusions about what parts of the data visualization in the Reflective phase the user was actually focused on. More recent developments in eye-tracking inside the HMD and foveated rendering might lead to the availability of advanced telemetry data so that researchers can derive more detailed information about a user's gaze than simplistic head orientation values.

### 4.4 Next steps

In further studies, we aim to explore a variety of potential adjustments for the Reflective phase by testing more interaction types included in the Reflective phase. For example, rather than presenting users with premade visualizations with minimal possible adjustments, users could **create their own** visualizations based on available data records, add **annotations**, or **compare** their own data with someone else's side by side via **arranged and coordinated views**. Also, the tasks for these studies are rather abstract and thus might resemble real-world VR training and coaching tasks only superficially. As a result, it could be interesting to design a **real-world user study** with professionals from an application domain (such as the medical or engineering fields).

### Author Contributions

Andreas Bueckle served as lead author, developed the VR applications for both user studies, and performed the data analysis for the Luddy VR user study, for which he also constructed the research design. Kilian Buehling and Andreas Bueckle compiled the study design for the RUI VR study.

Kilian Buehling performed the data analysis for the RUI VR study. Katy Börner and Patrick C. Shih assisted in developing the research design for both studies. Katy Börner provided substantial feedback on this manuscript throughout the process and oversaw Andreas Bueckle's work on this project as part of his dissertation.

## Funding

This research was partially funded by the National Institutes of Health (NIH) under grant OT2OD026671, the National Institute of Diabetes and Digestive and Kidney Diseases (NIDDK) Kidney Precision Medicine Project grant U2CDK114886, and the Common Fund Data Ecosystem (CFDE) OT2 OD030545. This project has also been funded in part with Federal funds from the National Institute of Allergy and Infectious Diseases (NIAID), National Institutes of Health, Department of Health and Human Services under BCBB Support Services Contract HHSN316201300006W/HHSN27200002 to MSC, Inc. [for an acknowledged contributor who is not an author on this publication]. The Stifterverband für die Deutsche Wissenschaft provided funding for Kilian Buehling's research visit at Indiana University, Bloomington via its INNcentive grant, which we gratefully acknowledge.

## Acknowledgments

We would like to thank Philip Beesley from Philip Beesley Architect Inc. for allowing us to use his 3D model of Luddy Hall at Indiana University; JangDong "JD" Seo from the Indiana Statistical Consulting Center for his expert input on the power analysis to determine the number of subjects needed to achieve significant results in our studies; Robert L. Goldstone for providing input on the initial research design for the RUI VR study; and Leonard Cross for his expert input during the development of the three RUI setups. We further express our gratitude to Kristen Browne (Bioinformatics and Computational Biosciences Branch, Office of Cyber Infrastructure and Computational Biology, National Institute of Allergy and Infectious Diseases, National Institutes of Health) for creating the 3D model of the kidney used in this study; to Perla Brown for improving the figures; and to the eight CNS colleagues who served as pilot testers. These user studies were conducted under Indiana University IRB protocol numbers 1910331127 and 1911941428.

## *Supplementary Material*

### 1 Ethics Statement

The research described in this paper was approved by the Institutional Review Board at Indiana University under protocol numbers 1910331127 (RUI VR) and 1911941428 (Luddy VR). Study information sheets for both studies can be found in a GitHub repository at https://github.com/cns-iu/optimizing-performance-in-VR-using-DVL-FW/tree/main/irb.

### 2 Supplementary Text for RUI VR

This GitHub repository contains relevant materials that did not make it into the paper given constraints around word count and the number of figures and tables allowed: https://github.com/cns-iu/optimizing-performance-in-VR-using-DVL-FW

It also contains C# scripts showing how we implemented interactions for the visualizations in the Reflective phases of the two VR studies.

### 2.1 Unity Projects

To ensure reproducibility of our results, and to enable all readers to replicate the steps in our user study, we made the Unity (Unity Technologies, 2021) projects for each study available on GitHub. Below are the links to each GitHub repository:

https://github.com/cns-iu/luddy_vr_unity_project
https://github.com/cns-iu/rui_vr_reflective_vr_project
https://github.com/cns-iu/rui_vr_user_study

### 2.2 Study design

The design for the RUI VR study is explained in more depth in an existing publication that focuses on comparing accuracy, completion time, and satisfaction between the three setups in the control cohort only. A paper describing the study design and results has been published recently (Bueckle et al., 2021a).

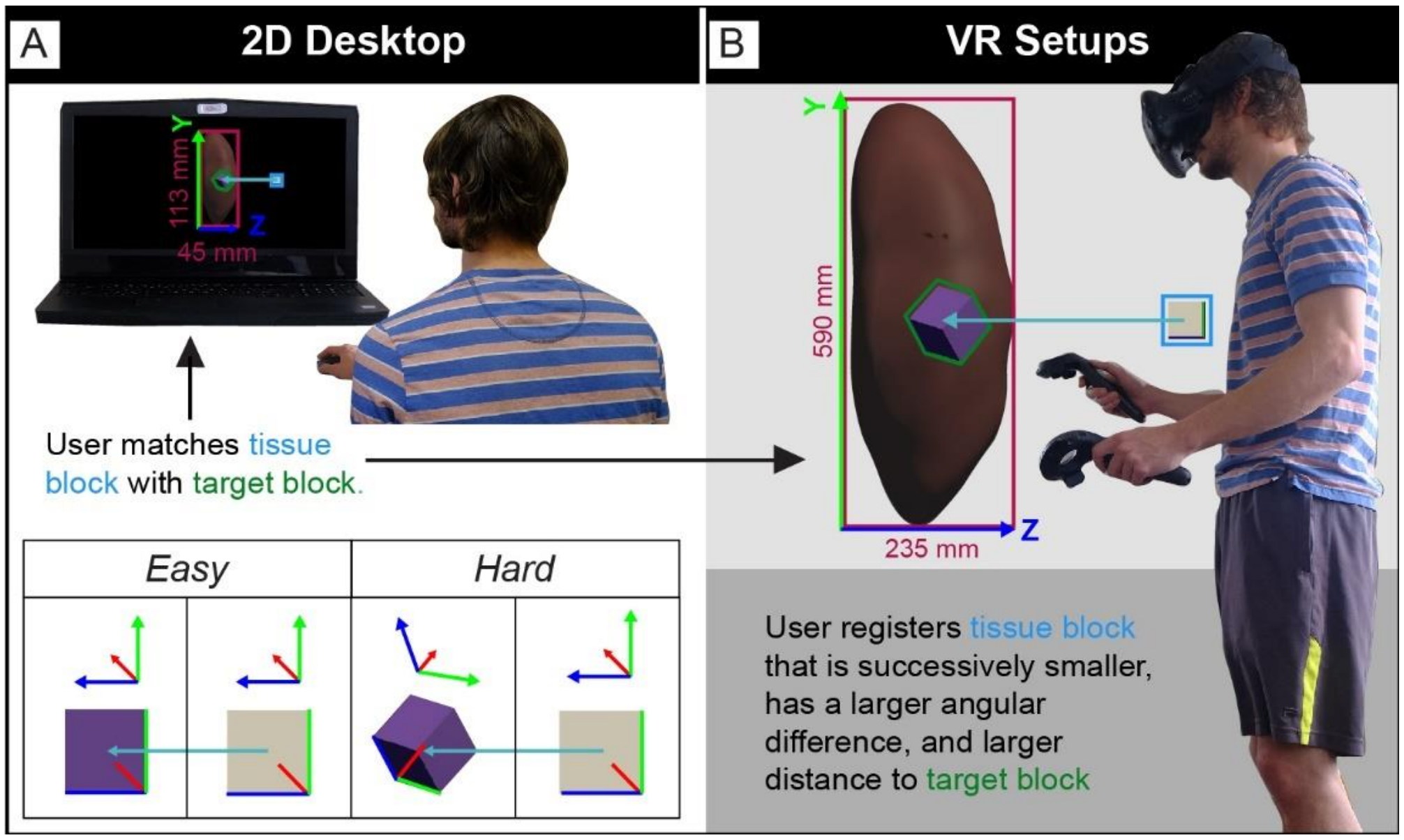

**Supplementary Figure 1.** The task setup in our user study: Reference organ with target block indicated (purple) and tissue block (white) to be registered into the target block. The light blue arrow indicates block centroid (mid-point) distance. Task difficulty increases as the tissue blocks get smaller, block rotation increases, and distance between the blocks increases. (A) 2D Desktop setup. (B) The two VR setups (Bueckle et al., 2021a).

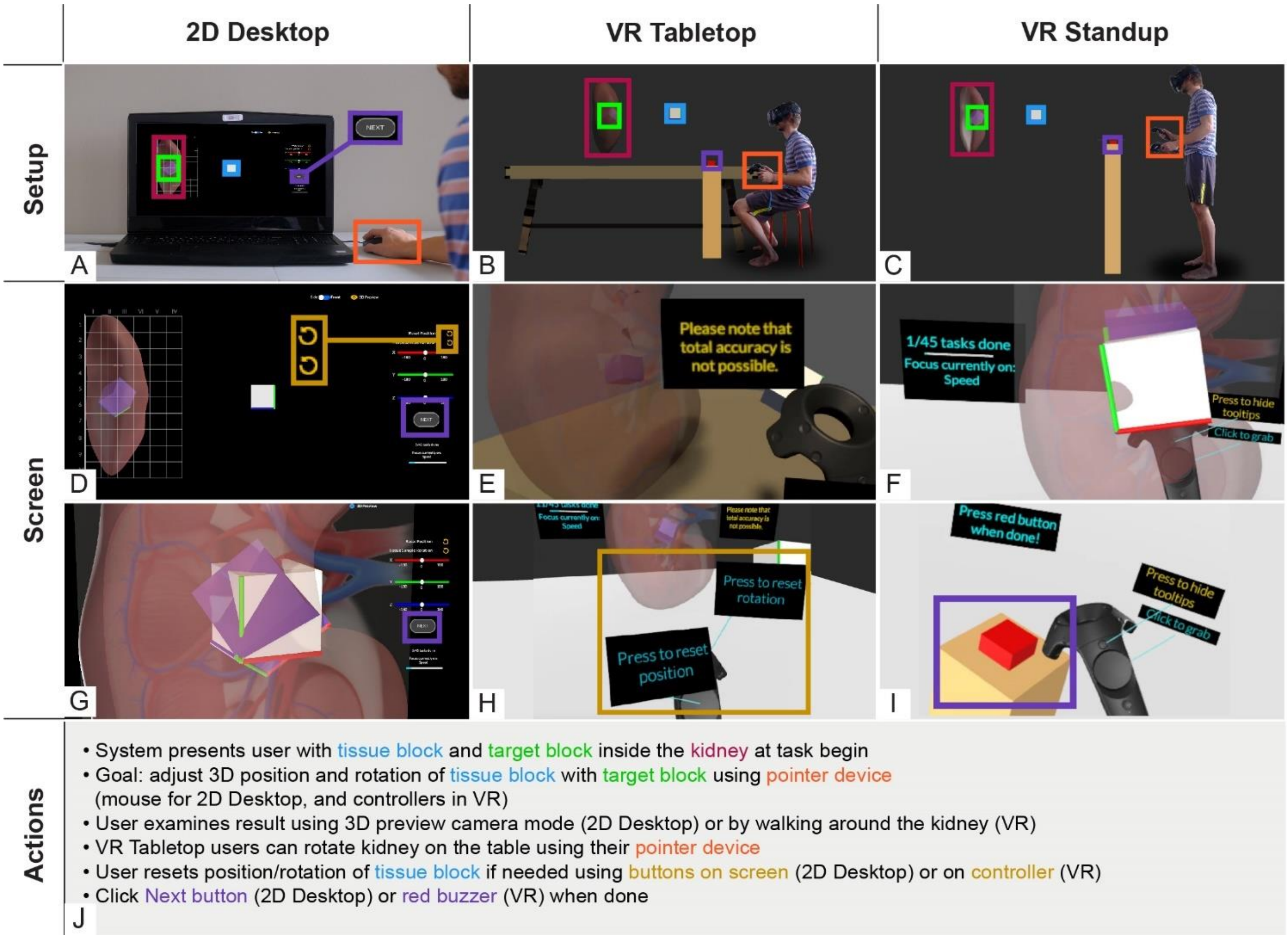

**Supplementary Figure 2.** Setup, screen, and actions for 2D Desktop, VR Tabletop, and VR Standup (Bueckle et al., 2021a). (A-C) Three RUI setups with a human subject. (D-I) screenshots of the user interface. (J) Required actions. The tissue block is outlined in blue, the target block in green, and the kidney—providing context and domain relevance—in pink. Tasks are submitted by selecting the purple NEXT/red button. The user could reset the position or rotation of the tissue block by pressing the corresponding yellow-brown virtual (2D Desktop) and physical buttons (VR).

$$Lerp = \left\{ \begin{matrix} a, & ,if\ t \leq 0 \\ b, & ,if\ t \geq 1 \\ a + (b - a) * t & ,if\ 0 < t < 1 \end{matrix} \right\}$$

**Supplementary Equation 1.** Formula to compute task difficulty.

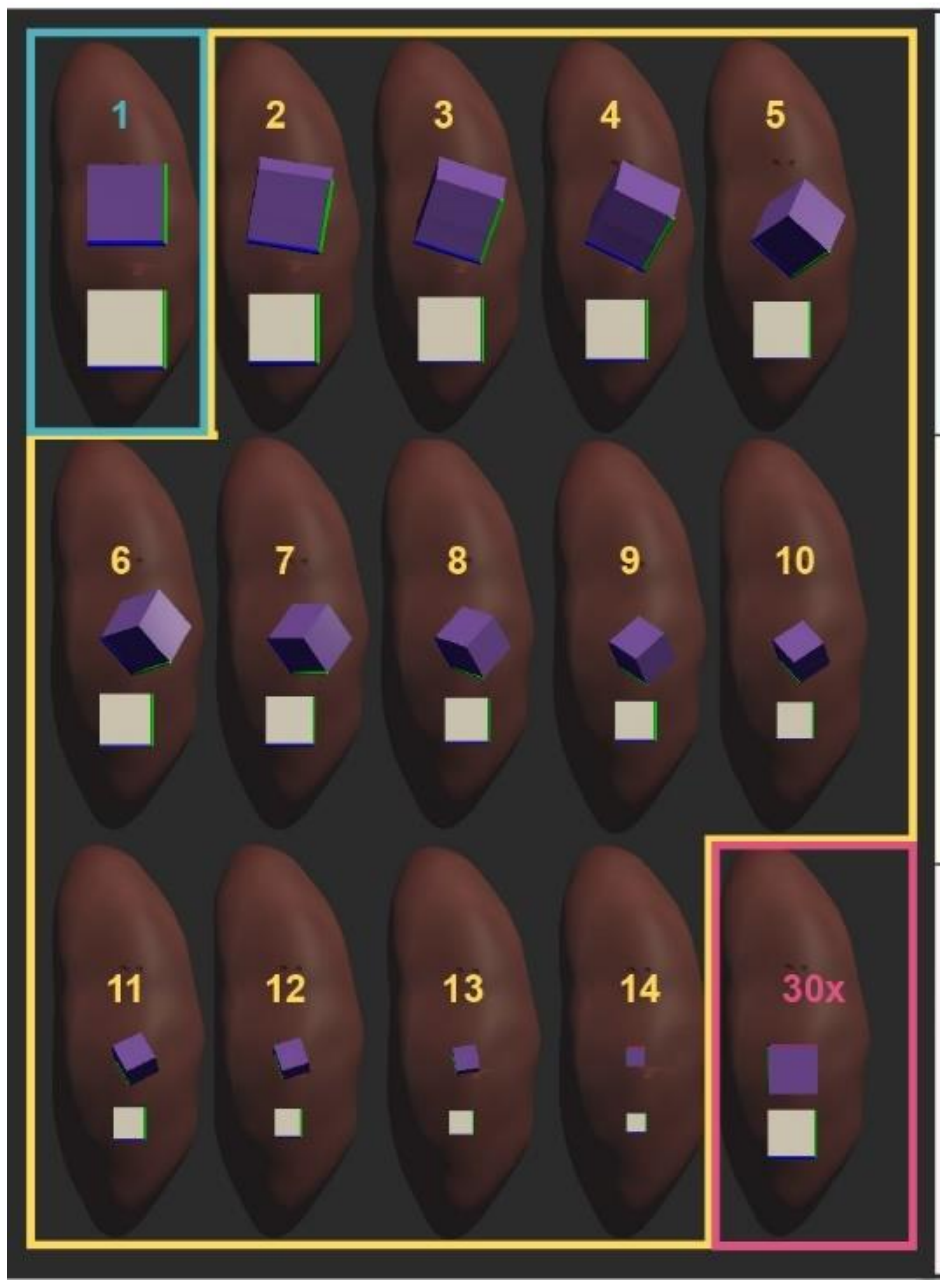

| | Distance | Angular difference | Size difference | Number of tasks | Prompt |
|---|---|---|---|---|---|
| Tutorial | *Smallest* | | *Biggest* | 1 | Audio explanation of interface |
| | 0.3 * kidneyHeight + offset | 0 degrees + offset | 0.2 * kidney Height | | |
| Ramp-Up | *Increasing* | | *Decreasing* | 14 | Odd: speed Even: accuracy |
| | start: 0.3 * kidney Height + offset end: 2 * kidneyHeight | start: 0 degrees + offset end: 180 degrees | start: 0.2 * kidney Height + offset end: 0.05 * kidney Height | | |
| Plateau | *Consistent* | | *Consistent* | 30 | Speed |
| | 1.15 * kidneyHeight | 180 degrees | 0.125 * kidneyHeight | | |

**Supplementary Figure 3. Task setup and levels of difficulty used in this study**. The offset (computed via **Supplementary Equation 1 above**) is a value that is added to gradually increase the distance and angular difference between the two blocks, and that is used to gradually decrease the size of the two blocks (Bueckle et al., 2021a).

## 2.3 Analysis of tool usage during Reflective phase

**Supplementary RQ**: In the Reflective phase, how do users apply the interactive tools?
**Supplementary Ha**: Most users will use the time slider to scroll through around 10 times the time span of their dataset.
**Supplementary Hb**: The most selected location for the play head of the slider will be towards the very end of the timecode in the dataset.
**Supplementary Hc**: Users will spend the majority of time with the kidney turned on as the presence of a reference organ is highly useful to understand the data overlay. The kidney is turned on by default.

We recorded a variety of metrics from VR subjects in the experiment group during the Reflective phase. This phase consisted of two parts: an intro where the user explored the best-performing user's dataset (with an audio tutorial about the interactive tools and goals of this phase) and, subsequently, the main part where they explored their own data. Supplementary Table 1 gives the number of observations (N), mean, standard deviation (SD), median, min, and max value for a series of metrics from the intro as well as the main part of the Reflective phase for both VR setups together and separate. Supplementary Table 2 provides definitions of variables for user behavior from the Reflective phase.

On average, users spent **464.63 seconds** (VR Tabletop, **SD = 149.82 s**) and **396.76 s** (VR Standup, **SD = 268.8 s**) in the main part of the Reflective phase. Unsurprisingly, VR Tabletop users traveled less with their HMDs (**26.88 meters**, **SD = 12.49 m**) than VR Standup users (**41.96 m**, **SD = 19.92** **m**).  Notice, however, the large **range 114.22 m** between the subject with the **most right-hand movement** and the subject with the **least** (both in the VR Standup setup)**.** Likewise, the subject with the most **distance traveled for the HMD** in the VR Standup setup was measured at **81.41 m** (vs. **13.64 m** for the least traveled), yielding a **range of 67.77 m**.

The bigger freedom of movement probably also to VR Standup users rotating their heads more, with **8434.43 degrees** vs. **7086.07 degrees** for VR Tabletop, equaling around 26 and 20 theoretical complete head rotations, respectively. Further, on average, VR Tabletop users spent **46.9%** of their time in the main part of the Reflective phase without the kidney, compared to **32.2%** for VR Standup users, prompting us to **confirm Hc** (users will spend the majority of time with the kidney turned on). Because VR Tabletop users could rotate the kidney when completing their tasks, kidney rotations were shown in the Reflective phase, which is users may have had more of an incentive to leave the kidney visible. The average number of tasks simultaneously visible was **6.42** (VR Tabletop, **SD = 3.96**) and **6.01** (VR Standup, **SD = 4.16**). Finally, with regards to time slider usage: VR Tabletop users moved the slider **almost twice as much on average** as VR Tabletop users. Specifically, VR Tabletop users scrolled through **5.15 times** the time span of their dataset, compared to **2.68 times** for VR Standup. Since both of these values are far off from the 10 times we predicted in **Ha**, we need to **reject Ha**.

Likewise, the mean position of the raw slider on a scale from 0 (first time stamp, beginning of the dataset) to 1 (last time stamp, end of the dataset) was similar for both setups at **0.79** for VR Tabletop and **0.73** for VR Standup users, requiring us to **reject Hb** (predicting that the most selected location for the slider would be towards the very end of the dataset).

## 2.4 Metrics

To analyze survey and task data, we defined three performance metrics (position accuracy, rotation accuracy, and completion time) as well as a satisfaction score.

### 2.4.1 3D position accuracy

We defined position accuracy as the distance of the centroids of the tissue block and the target block, see light blue arrow in Supplementary Figure 1. We compute the distance at run time using `Vector3.Distance()`, a static method in Unity that returns the distance between two points in 3D space. The position of both blocks and the centroid distance was collected at 10 Hz (i.e., 10 times each second).

To make use of the various possibilities for scaling in VR, the kidney was displayed in different heights across setups (but always with the same width-to-height-to-depth ratio). Measured from the lowest to the topmost vertex, the kidney in the two VR setups was 0.59 Unity scene units tall. In VR, scene units correspond to physical meters, so the kidney appeared at a height of 590 mm. Similarly, in the 2D Desktop setup, the kidney appeared at a height of 113 mm on the laptop display (see Supplementary Figure 1). In order to compare position accuracy results between 2D Desktop and the VR setups, we normalized these values by dividing them by the height in which the kidney appeared to the user.

### 2.4.2 3D rotation accuracy

Rotation accuracy equals the angular difference between the two tissue blocks at task submission (see Supplementary Figure 1). For ease of analysis, it was reduced to an individual number between 0 (exact same rotation) and 180 (diametrically opposite rotation). We used Unity's built-in Quaternion.Angle() function to compute this angle. Angle() takes two orientations, each consisting of three angles, expressed either as Euler angles or Quaternions, and returns a single float value between 0 and 180.

This means that several combinations of different rotations between tissue block and target block could yield the same angular difference. In order to preserve as much detail about the subject's action as possible, equivalent to the position, we logged the rotation of both blocks throughout the experiment as well.

#### 2.4.3 Completion time

Completion time refers to the amount of time between the submission of a task and the submission of the previous task. Completion time is measured in seconds.

#### 2.4.4 Performance plateau

During the Plateau phase (**Supplementary Figure 3**), subjects performed 30 identical tasks, providing a unique opportunity to identify if and when a subject achieves a performance plateau. A plateau of a performance variable (task completion time, centroid accuracy, or rotation accuracy) is reached when the deviation of the performance variable does not exceed the mean performance of the subject until the end of the Plateau phase. As mean performance, we consider the average performance in a moving window of 20 tasks of the subject to reduce the influence of possible performance outliers. This width of the moving window supplies a stable mean by considering a certain inertia in performance improvement without including at all times the extreme values that can often be found towards the beginning and the end of the Plateau phase. For each subject, we analyzed after which task the performance stabilized by iterating through a recursive process, in which the relative deviation of the last task of the Plateau phase is calculated. If it does not exceed one (thus if the deviation of the performance variable in this task is not higher than its mean) we iterate this calculation for the previous task until we arrive at a task where the relative deviation is larger than 1. We consider all tasks after this (until the last task of the phase) to be on a performance plateau. If a subject reaches a performance plateau, we take the average performance (for example, mean completion time per task) of all the tasks that are completed after reaching this plateau.

#### 2.4.5 Satisfaction

To assess user satisfaction, we included a corresponding item in the post-questionnaire via a five-point Likert scale from one (not at all satisfied) to five (very much satisfied), with three being a neutral value, and we report results aggregated by setup.

## 3 Data Availability

We made the raw data generated for our paper available via Zenodo (Bueckle et al., 2021c; b). More information can be found in our GitHub repository for Supplemental Material at https://github.com/cns-iu/optimizing-performance-in-VR-using-DVL-FW#raw-data.

## 4 Supplementary Figures

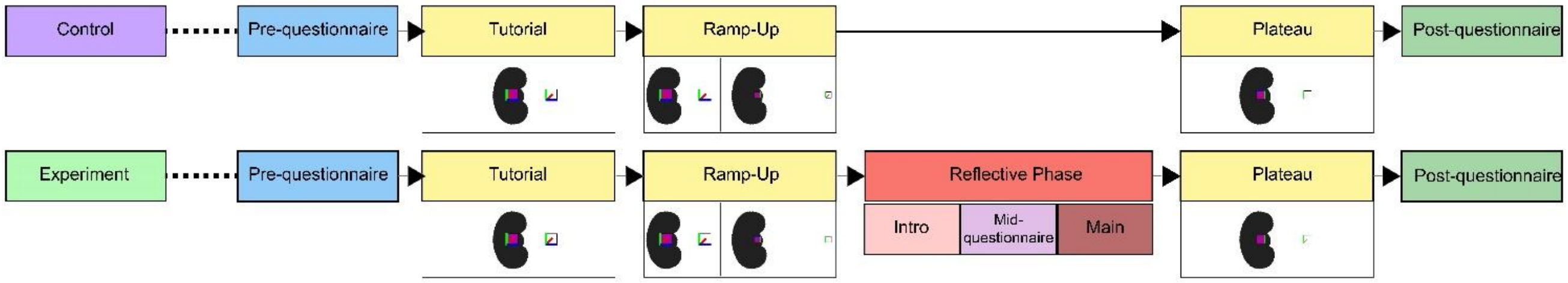

**Supplementary Figure 4.** The study design of the RUI VR user study in detail. Both cohorts filled out a pre-questionnaire before completing a series of VR tasks in three stages: Tutorial (one task, not timed or counted towards performance), Ramp-Up (14 increasingly difficult tasks), and Plateau (30 identical tasks). The experiment cohort participated in a Reflective phase, separated into an Intro part (where they investigated a visualization of the tissue block placements and head/hand movements of the highest performing user in their setup) and a Main part (where they investigated the data from their own Ramp-Up phase. Note that we did not explicitly tell the experiment users that the data in the Intro part came from the highest-performing user. Finally, both cohorts filled out a post-questionnaire to assess satisfaction and gather feedback about the experiment and the setup that the subject was using to complete their tasks (2D Desktop, VR Tabletop, and VR Standup).

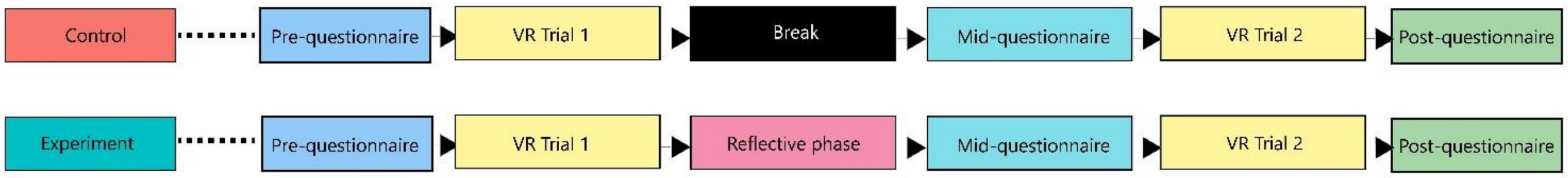

**Supplementary Figure 5.** The study design of the Luddy VR user study in detail. Both cohorts started with a pre-questionnaire before completing 24 navigation tasks (including 4 tutorial tasks) during VR Trial 1. The control cohort then took a break, and the experiment cohort investigated their own data in a Reflective phase. Both cohorts then filled out a mid-questionnaire about the virtual building, the number of tasks in total, per floor, etc. Finally, all subjects completed a post-questionnaire to assess satisfaction and gather feedback about the experiment.

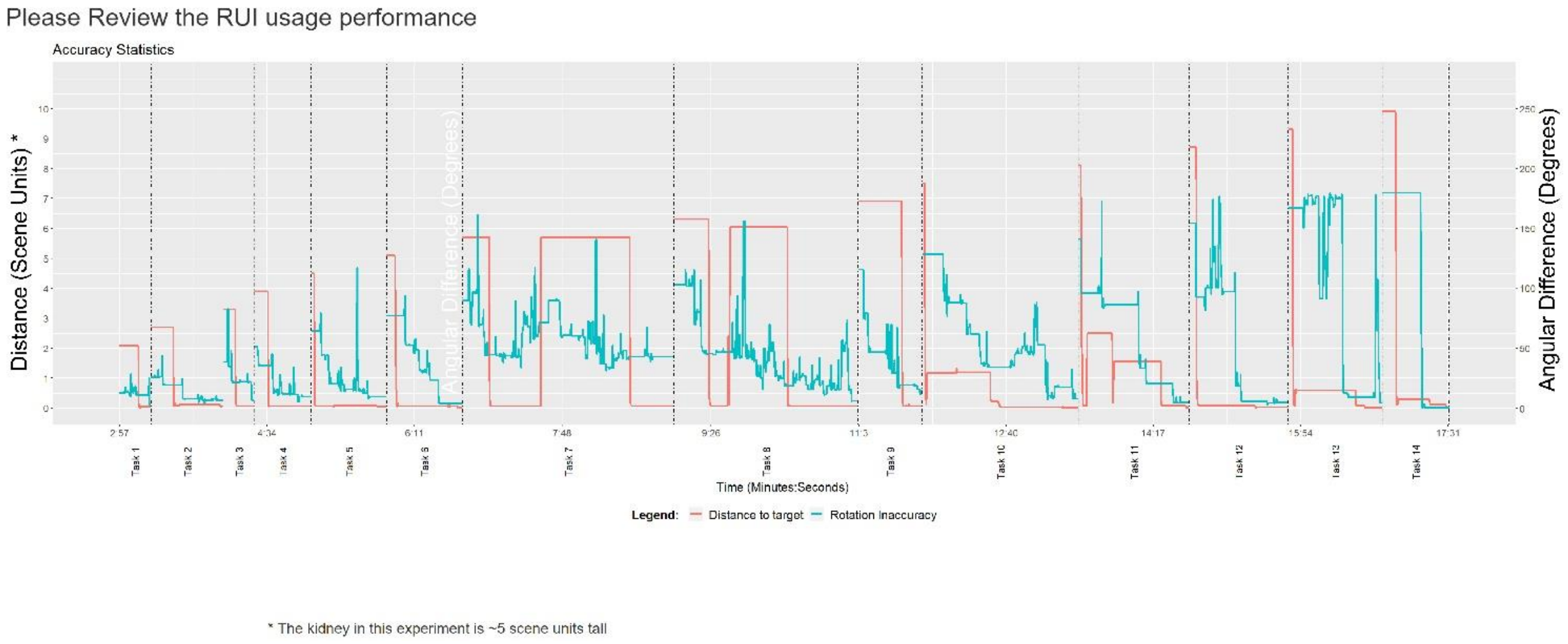

**Supplementary Figure 6.** Line graph of distance between tissue and target block (orange) and angular difference (green) for the best user in the control cohort for 2D Desktop.

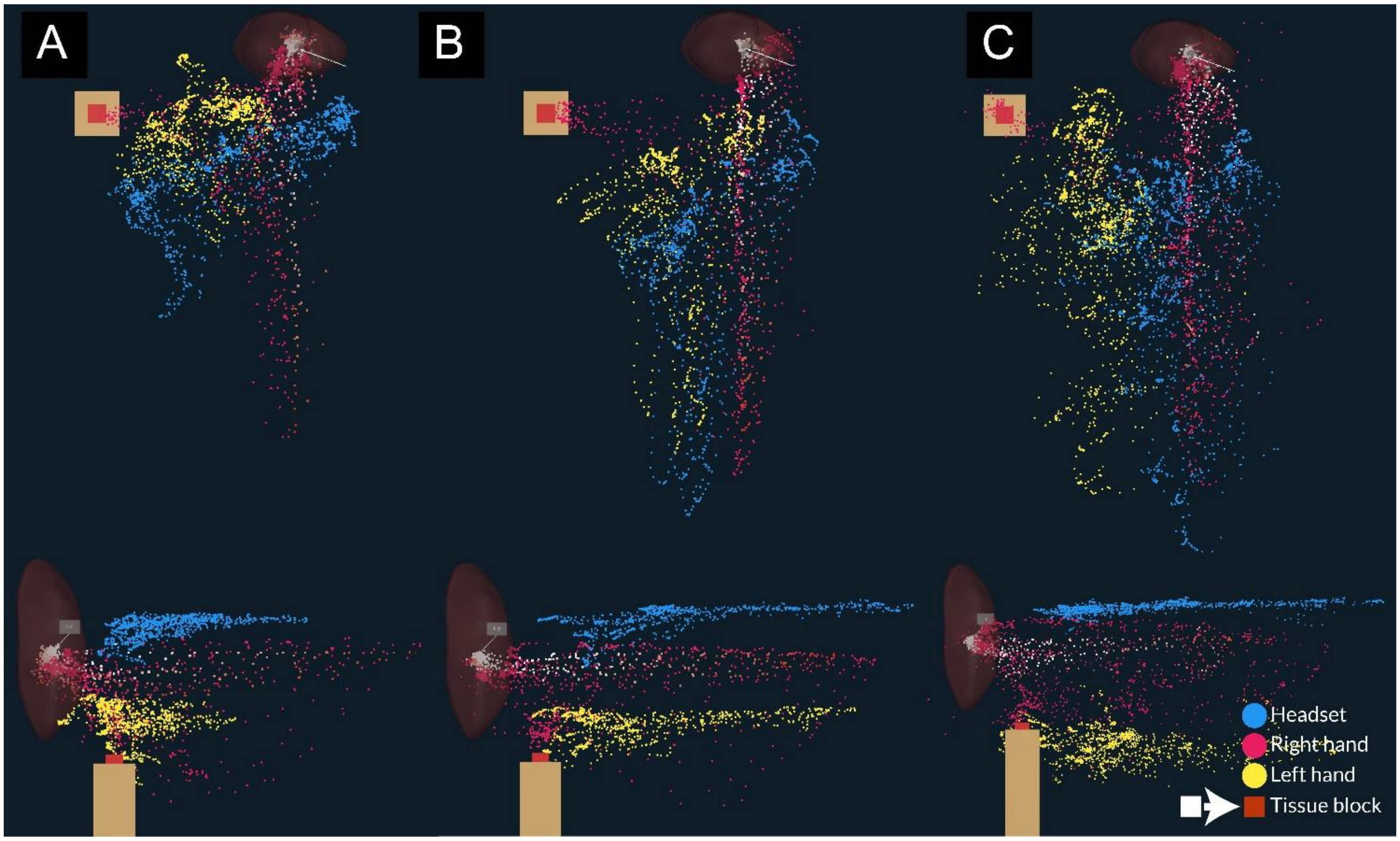

**Supplementary Figure 7.** Top view visualizations for three subjects with various spatial usage patterns (Standup). Left: concentrated work on one axis (see HMD in blue). Middle: plenty of back-and-forth along the z-axis. Right: wide spread around the z and x-axis.

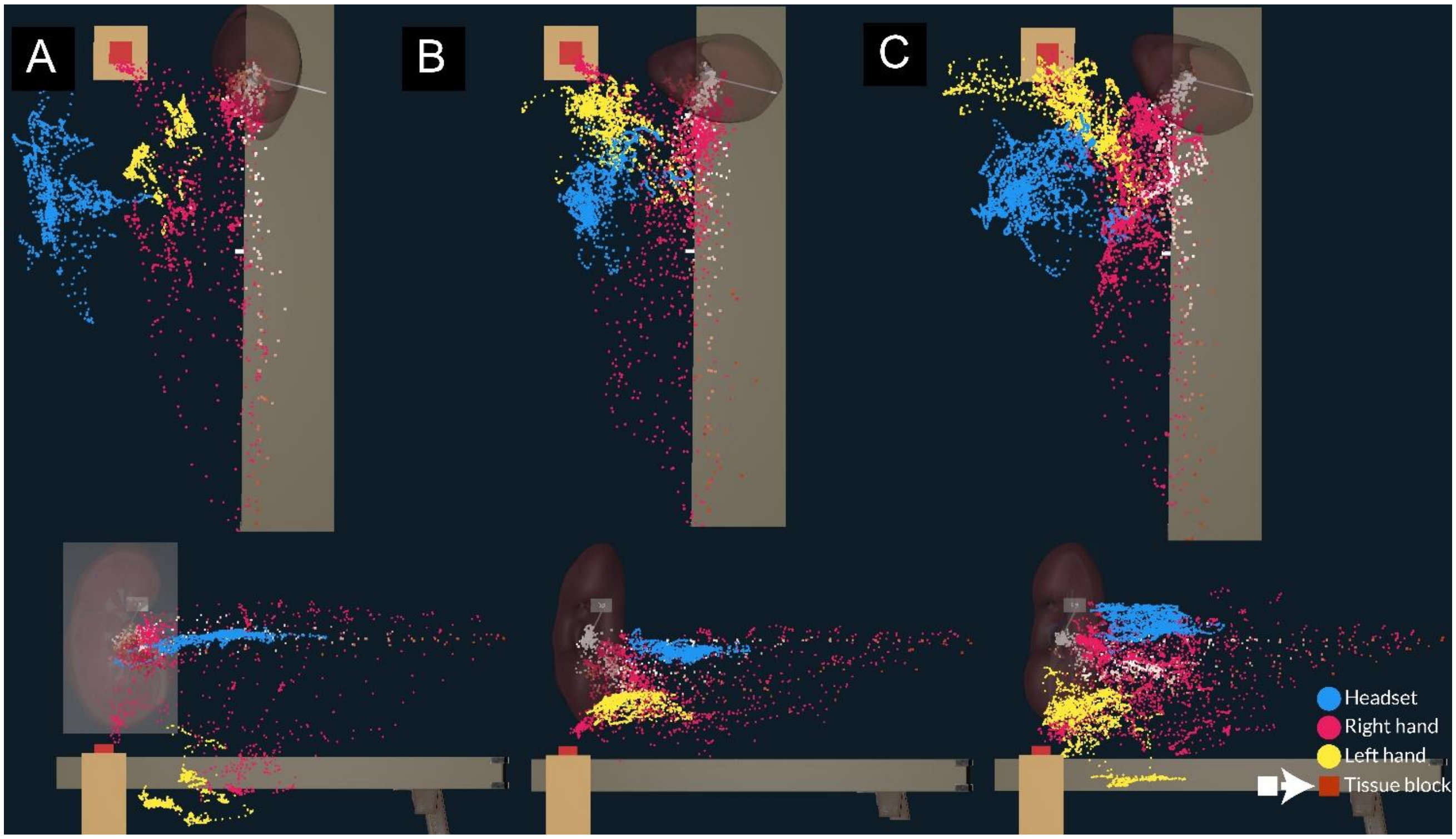

**Supplementary Figure 8.** Top view visualizations for three subjects with various spatial usage patterns (Tabletop). A: concentrated work on one axis. Middle: plenty of back-and-forth along the z-axis. Right: wide spread around the z and x-axis.

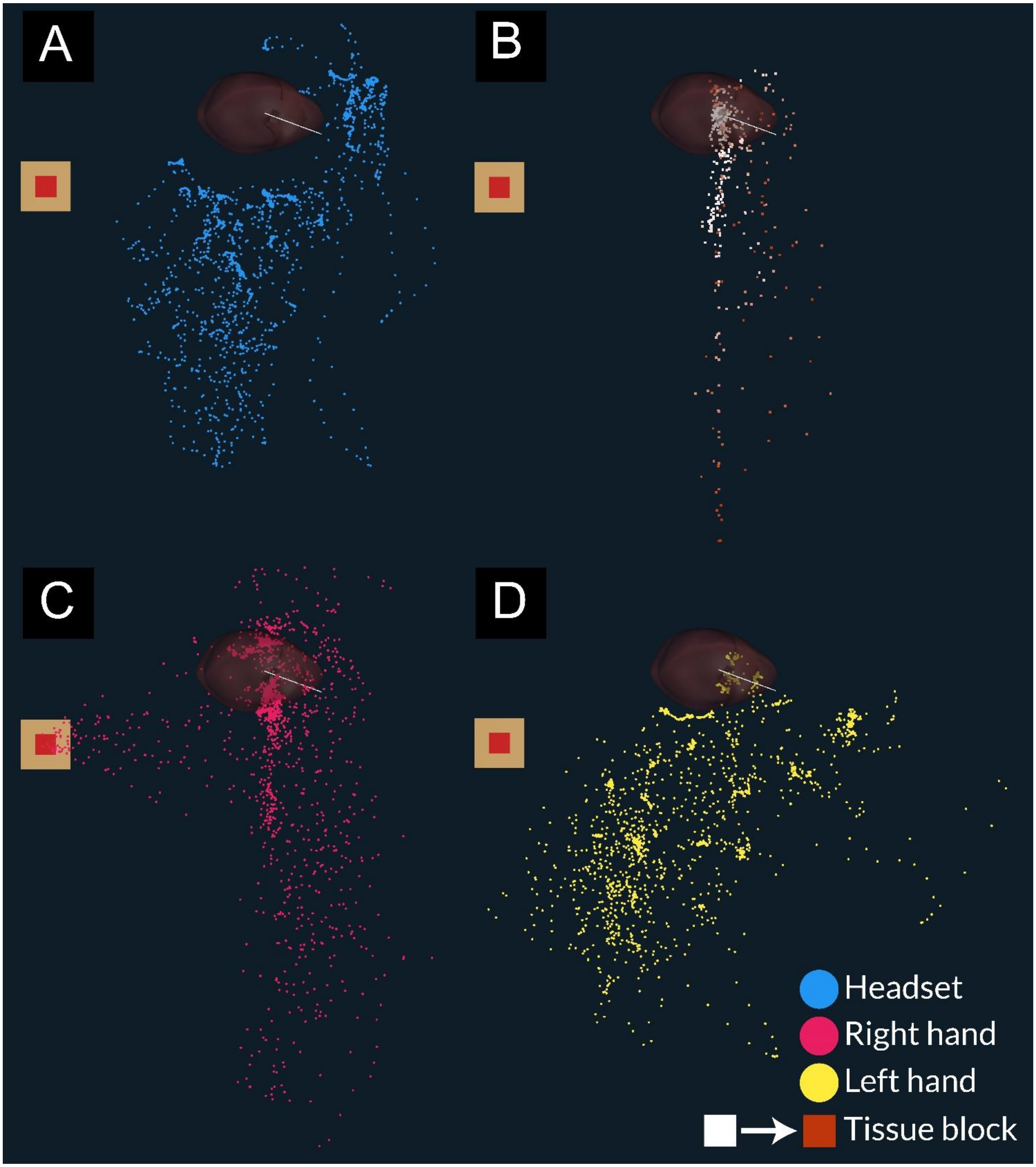

**Supplementary Figure 9.** Data overlay for the VR setups separated by color. A: HMD. B: tissue block. C: right controller. D: left controller.

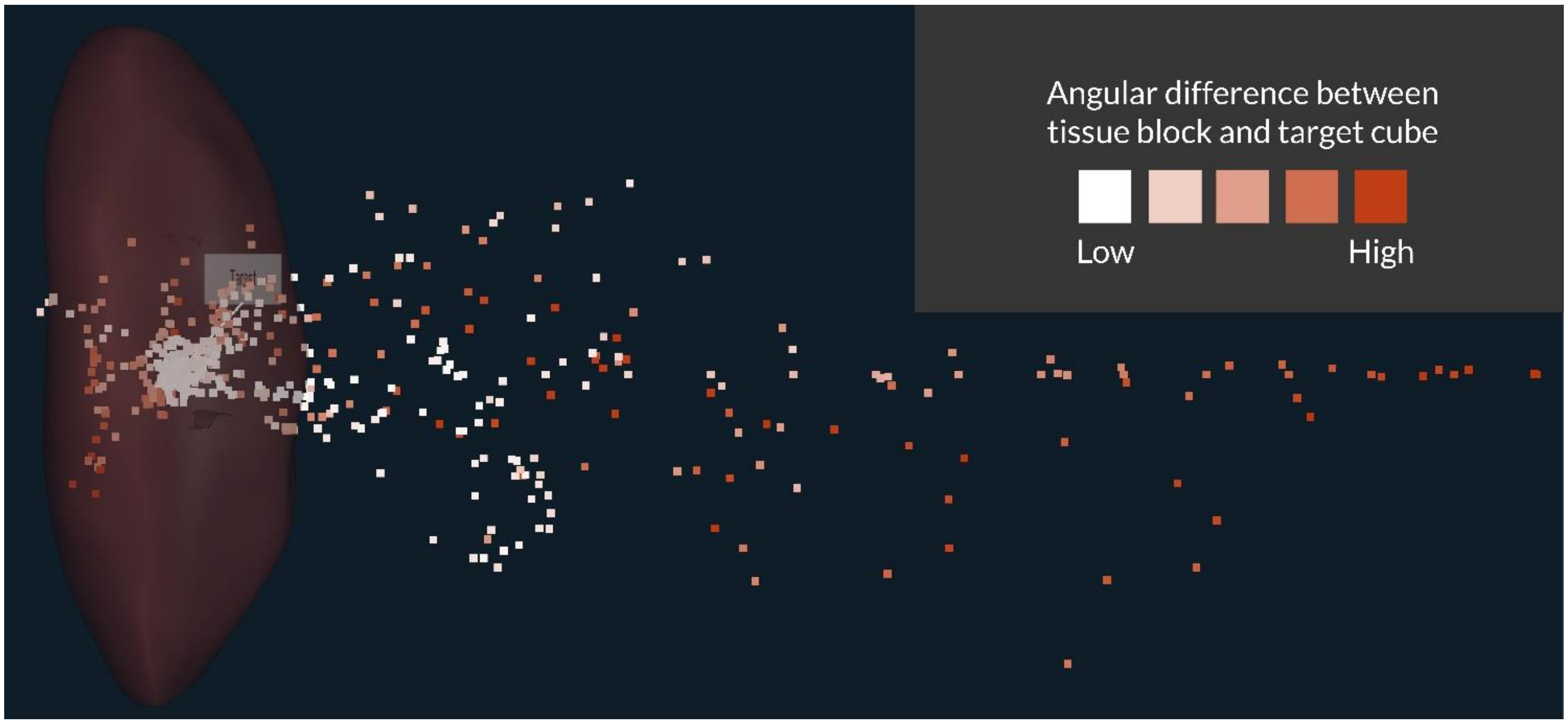

**Supplementary Figure 10.** Distribution of tissue block locations over time, with angular difference between the tissue and target block encoded with a sequential color scheme.

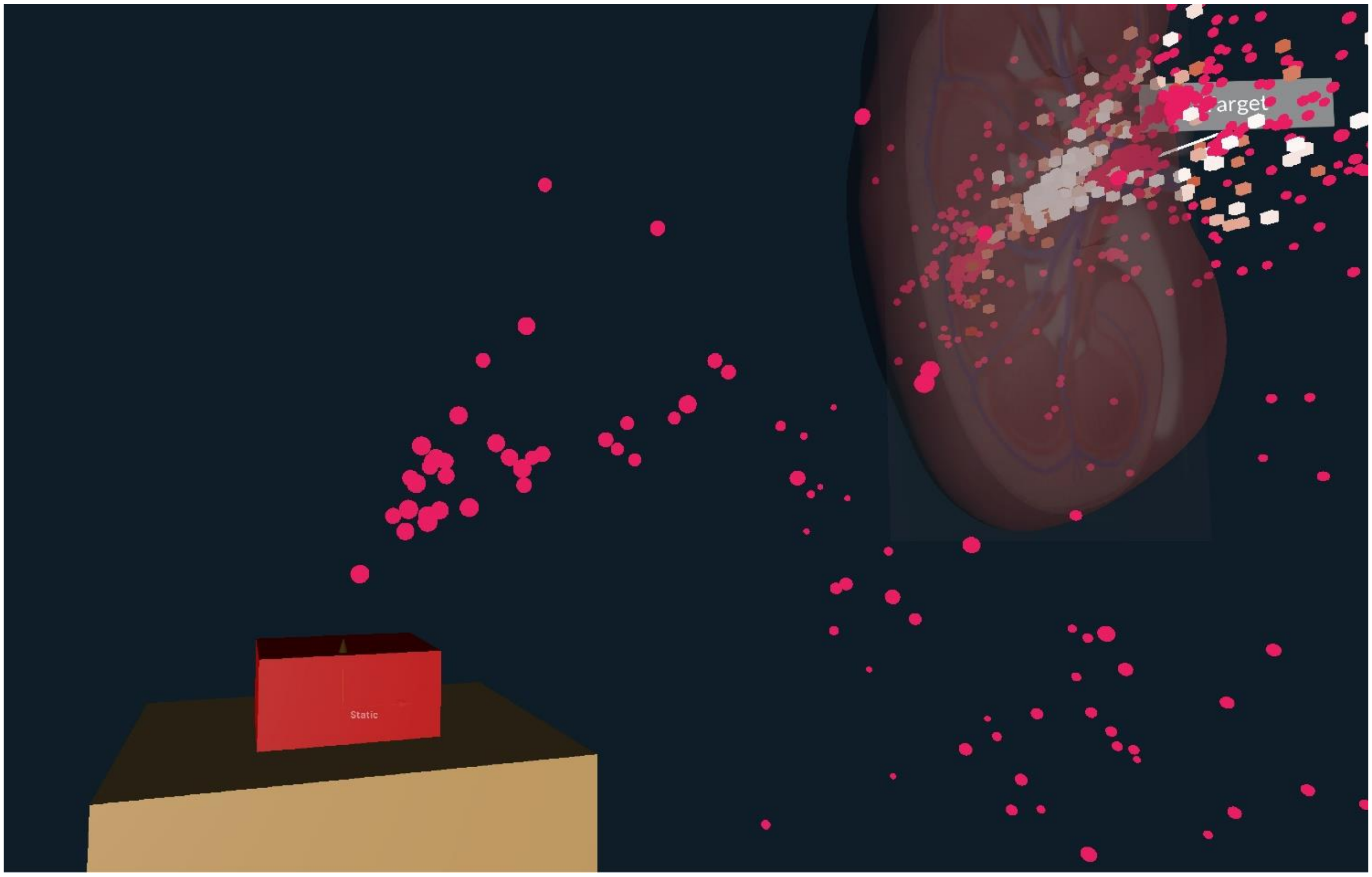

**Supplementary Figure 11.** The user's repeated pressing of the virtual red buzzer produces a hot spot.

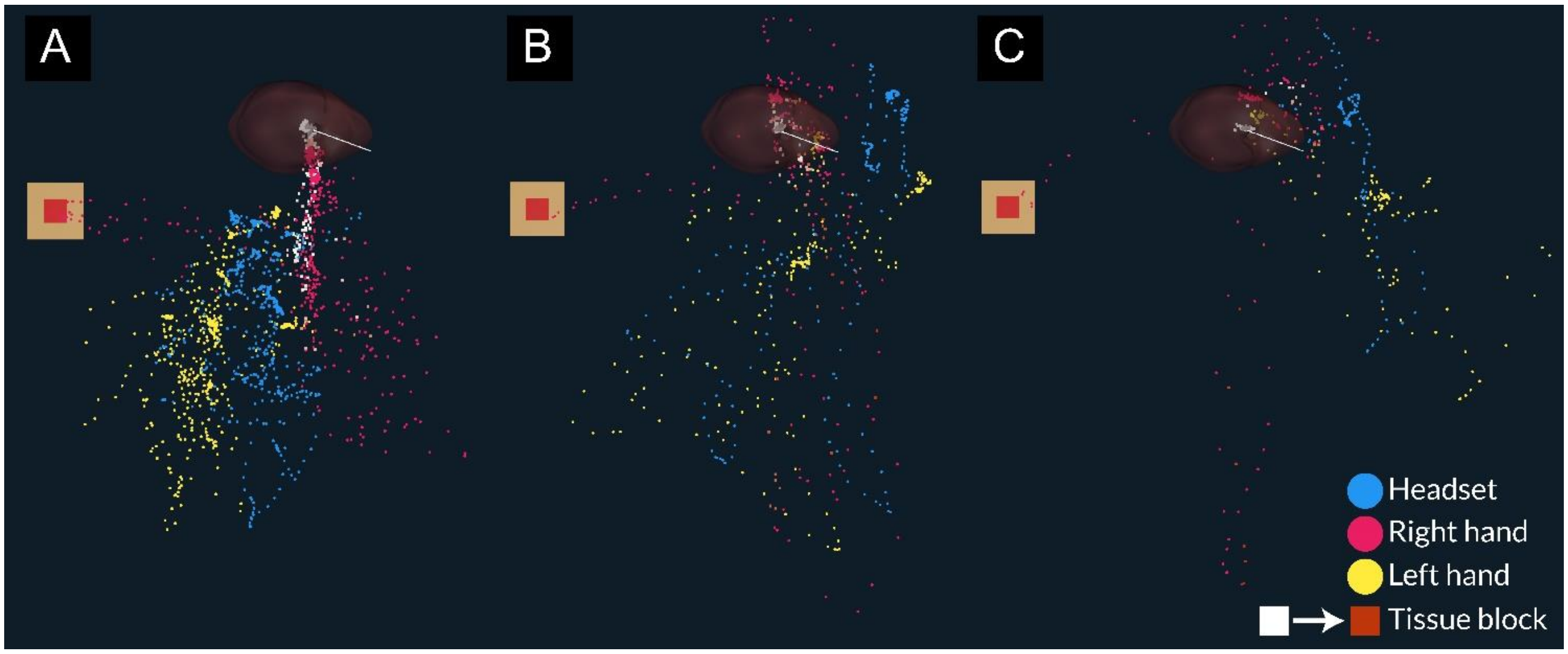

**Supplementary Figure 12.** VR Standup user with three different stages shown. A: Tasks 1-5. B: Tasks 10-11. C: Only task 14

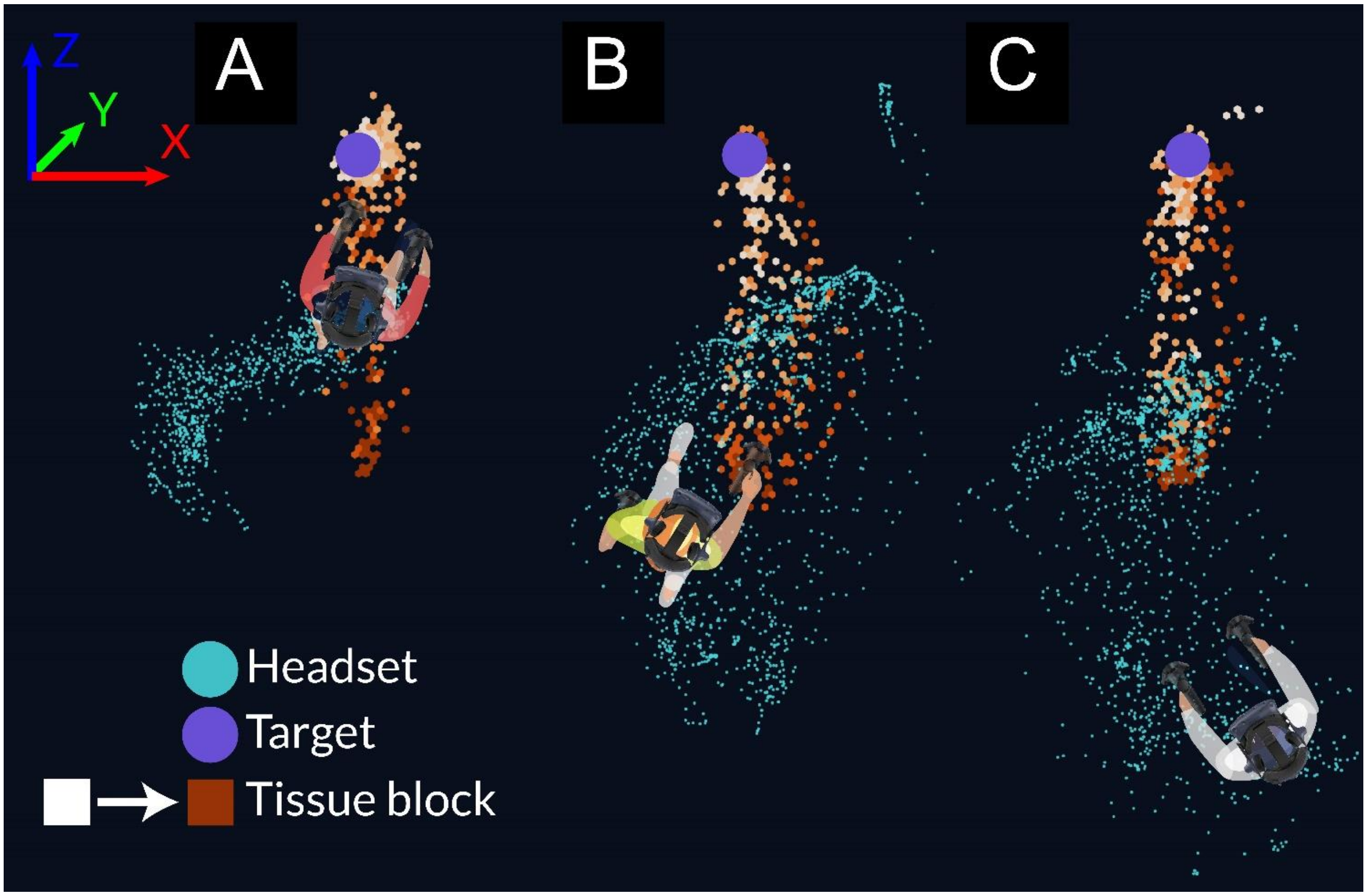

**Supplementary Figure 13**. Overhead view (x-z plane) of three participants (A, B, C) with unique movement patterns during the Ramp-Up phase.

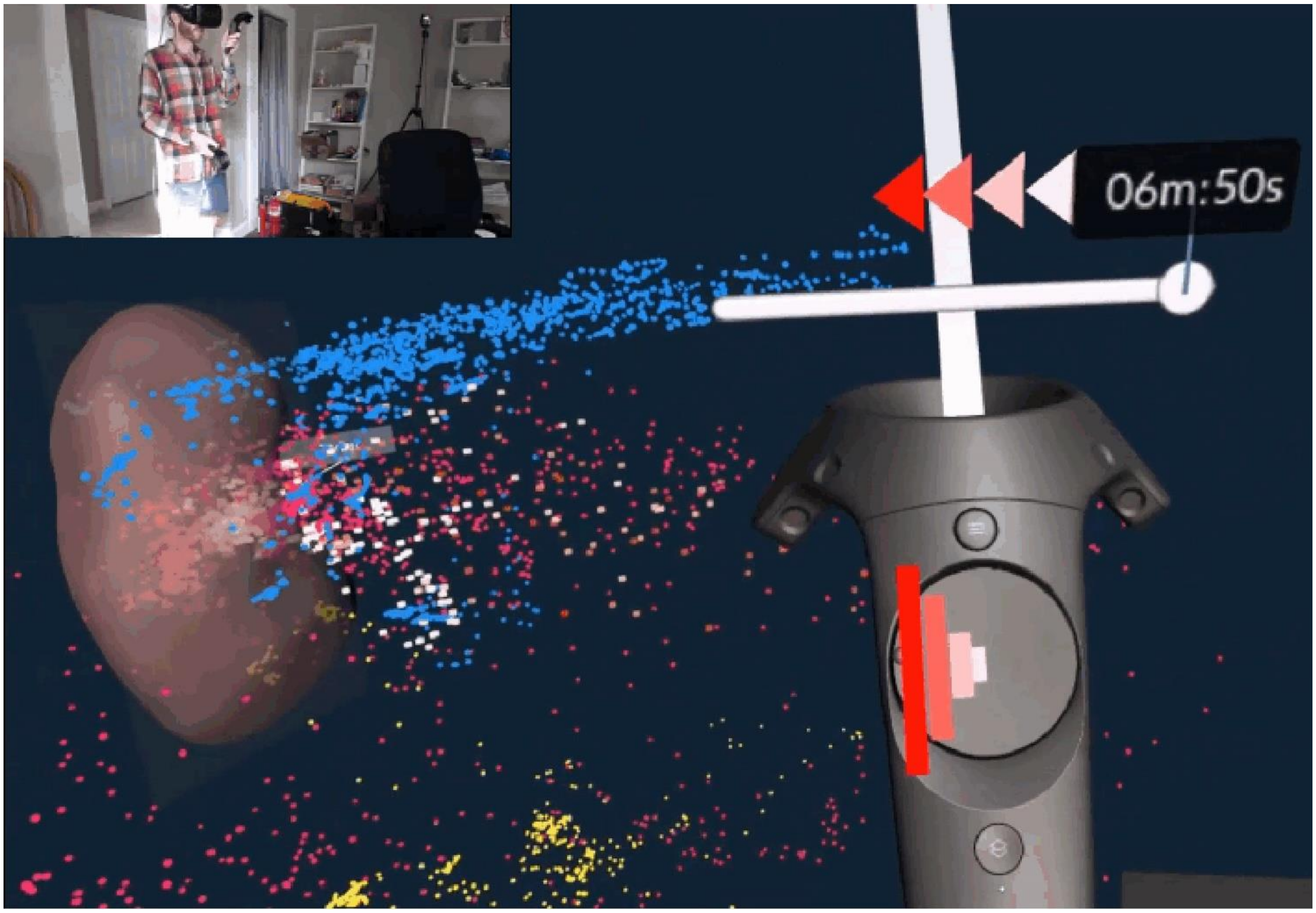

**Supplementary Figure 14.** Time slider to skip forward and backward in time by using via the thumbpad on the VR controller.

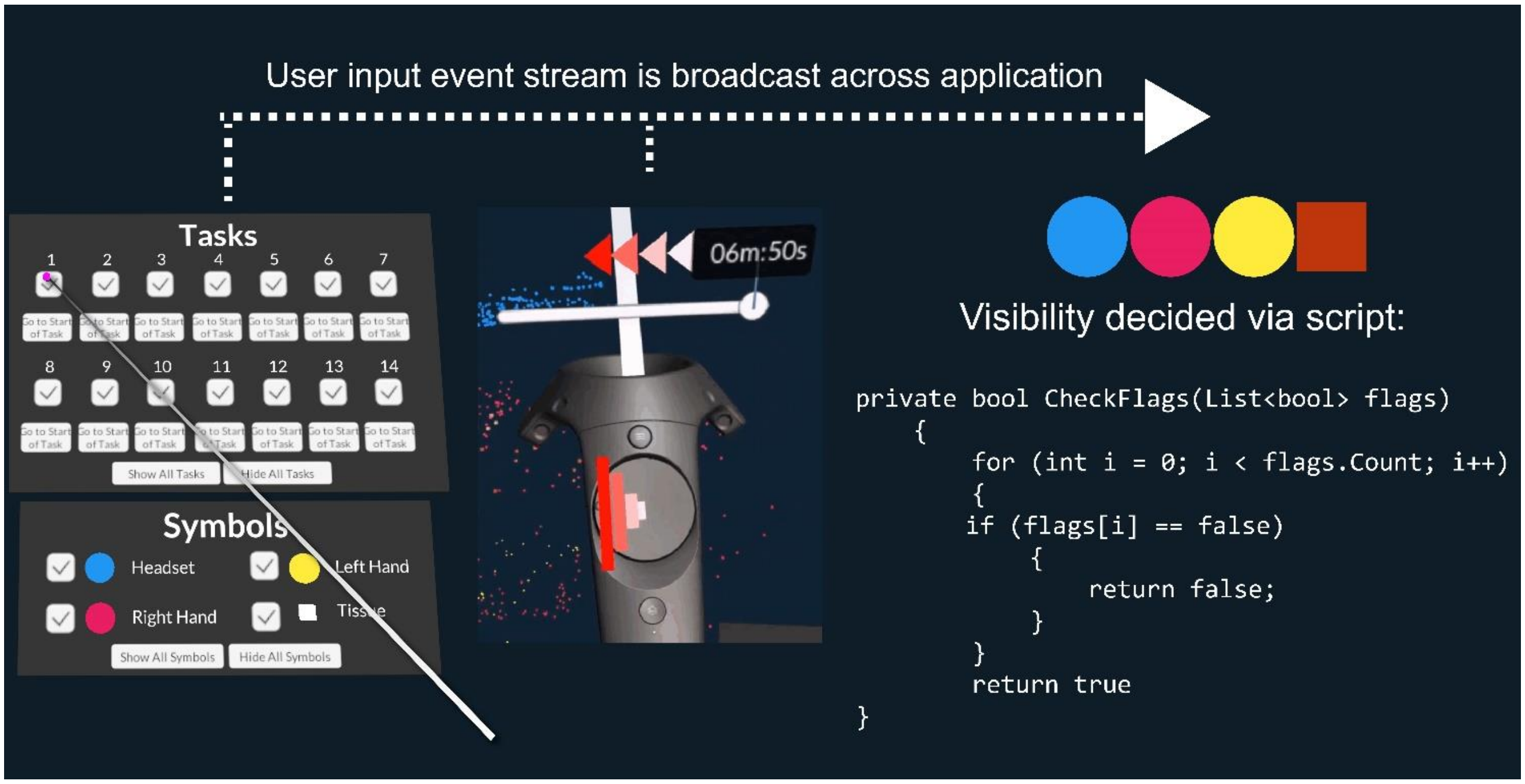

**Supplementary Figure 15.** Abstract image of the event listener for each graphic symbol. A series of Boolean values is passed into the CheckFlags() function, for example, whether the currently shown time stamp is later than the associated time stamp of the graphic symbol. If all arguments evaluated to true, the graphic symbol was displayed.

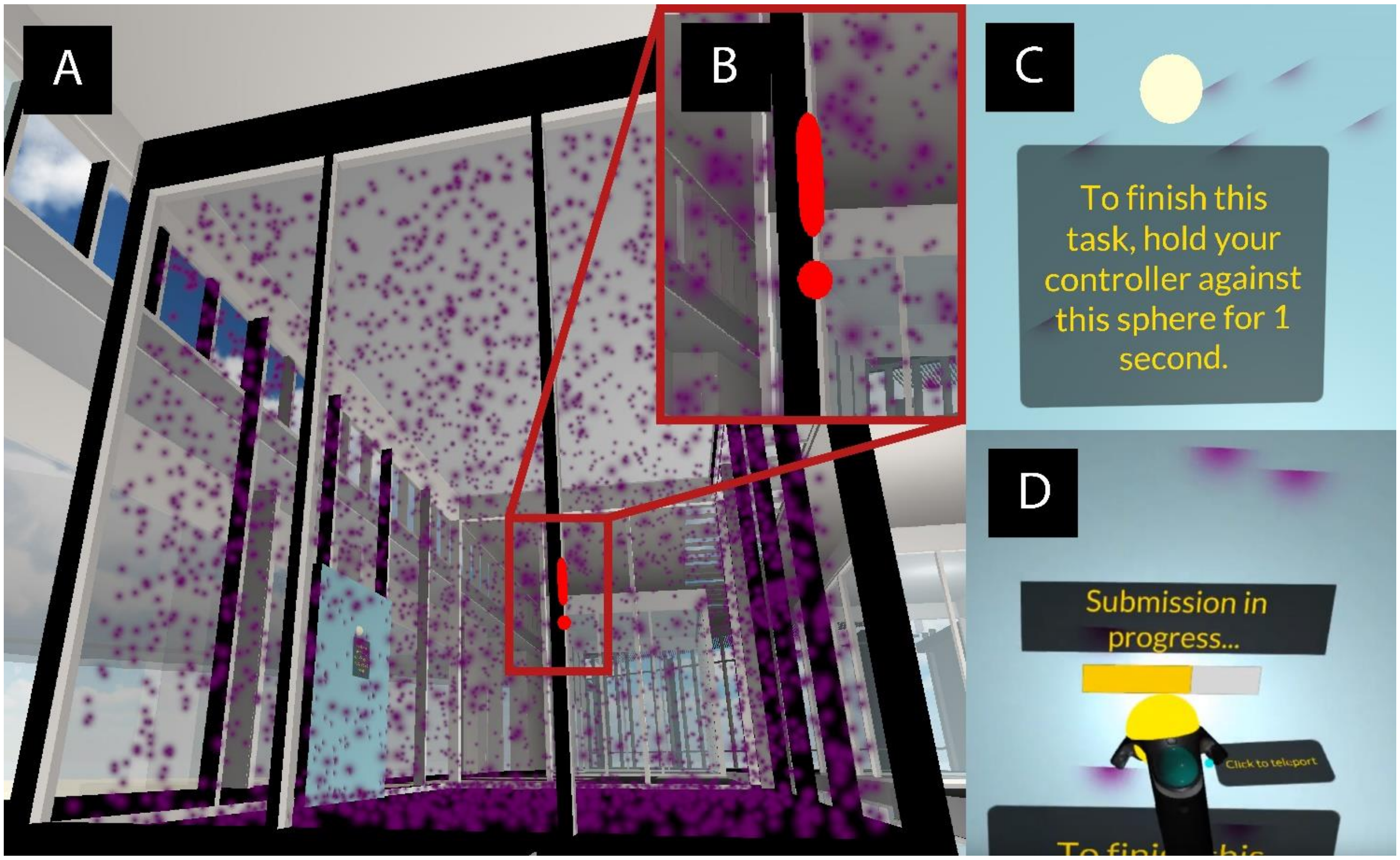

**Supplementary Figure 16.** Anatomy of a task room. A: the tutorial task room from the outside. B: note that the red exclamation mark is rendered on top of the wall in front of it. C: task submission instructions. D: the submission in progress.

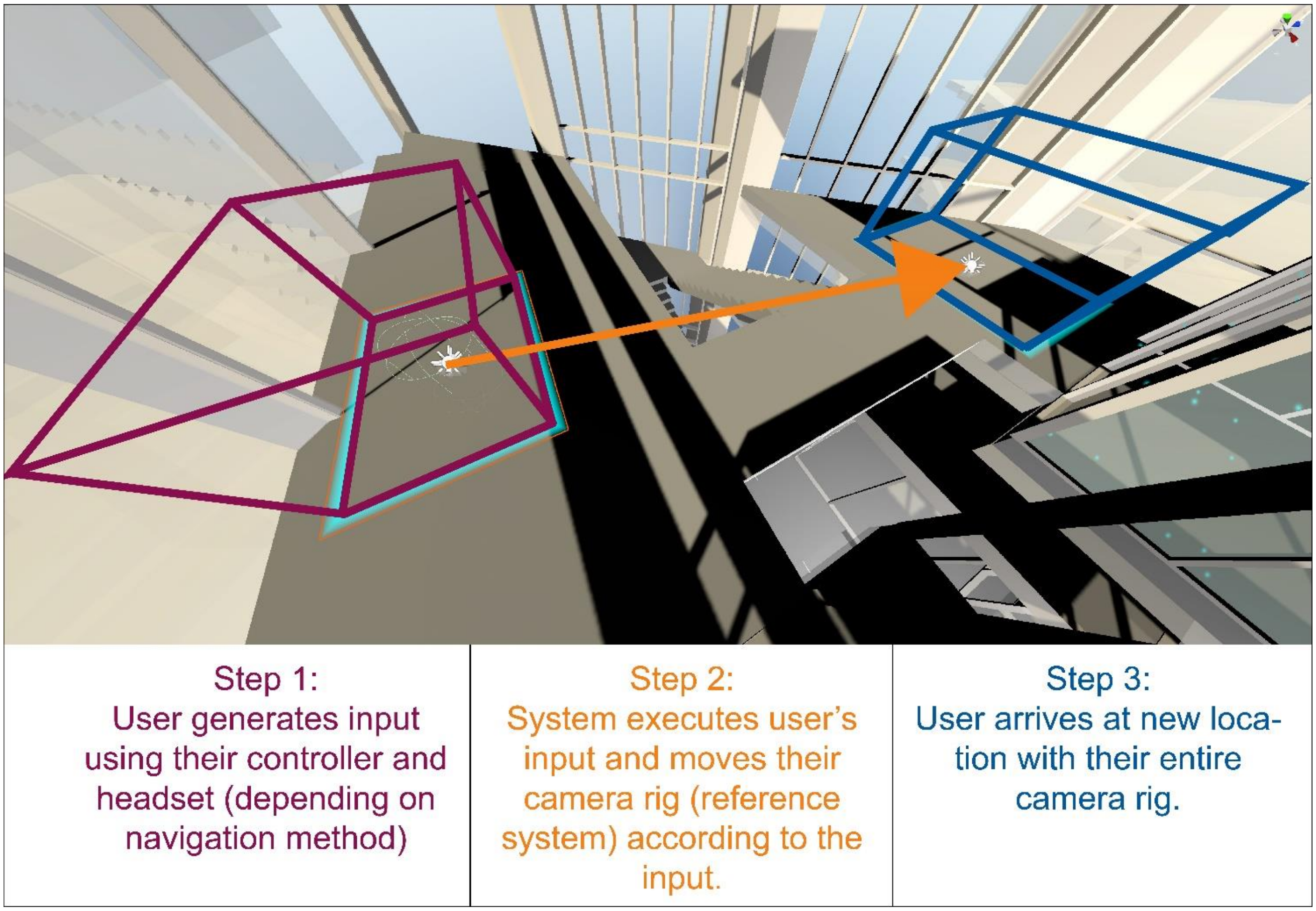

**Supplementary Figure 17.** Illustration of basic navigation.

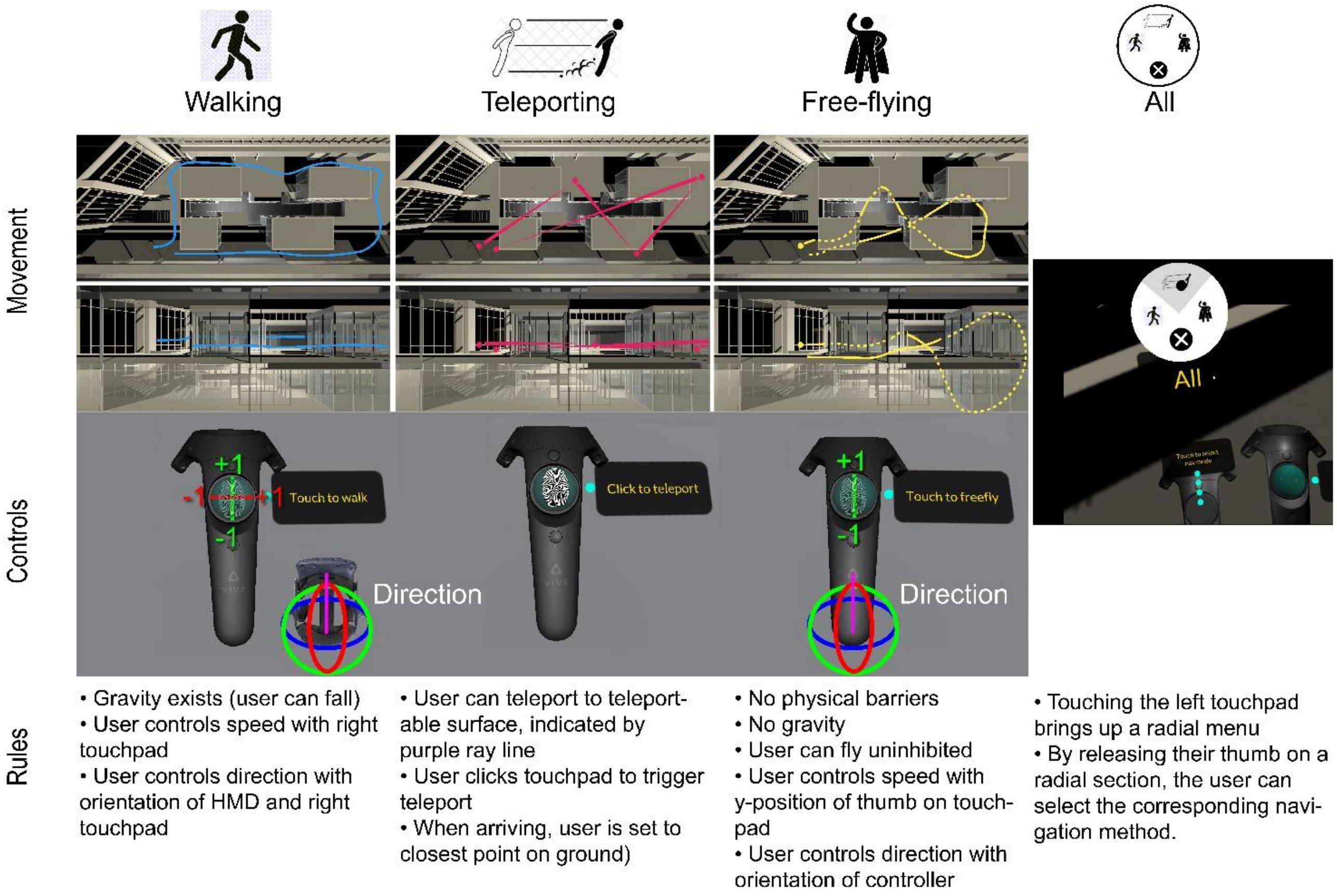

**Supplementary Figure 18.** A comparison of movement, controls, and features for all three navigation methods.

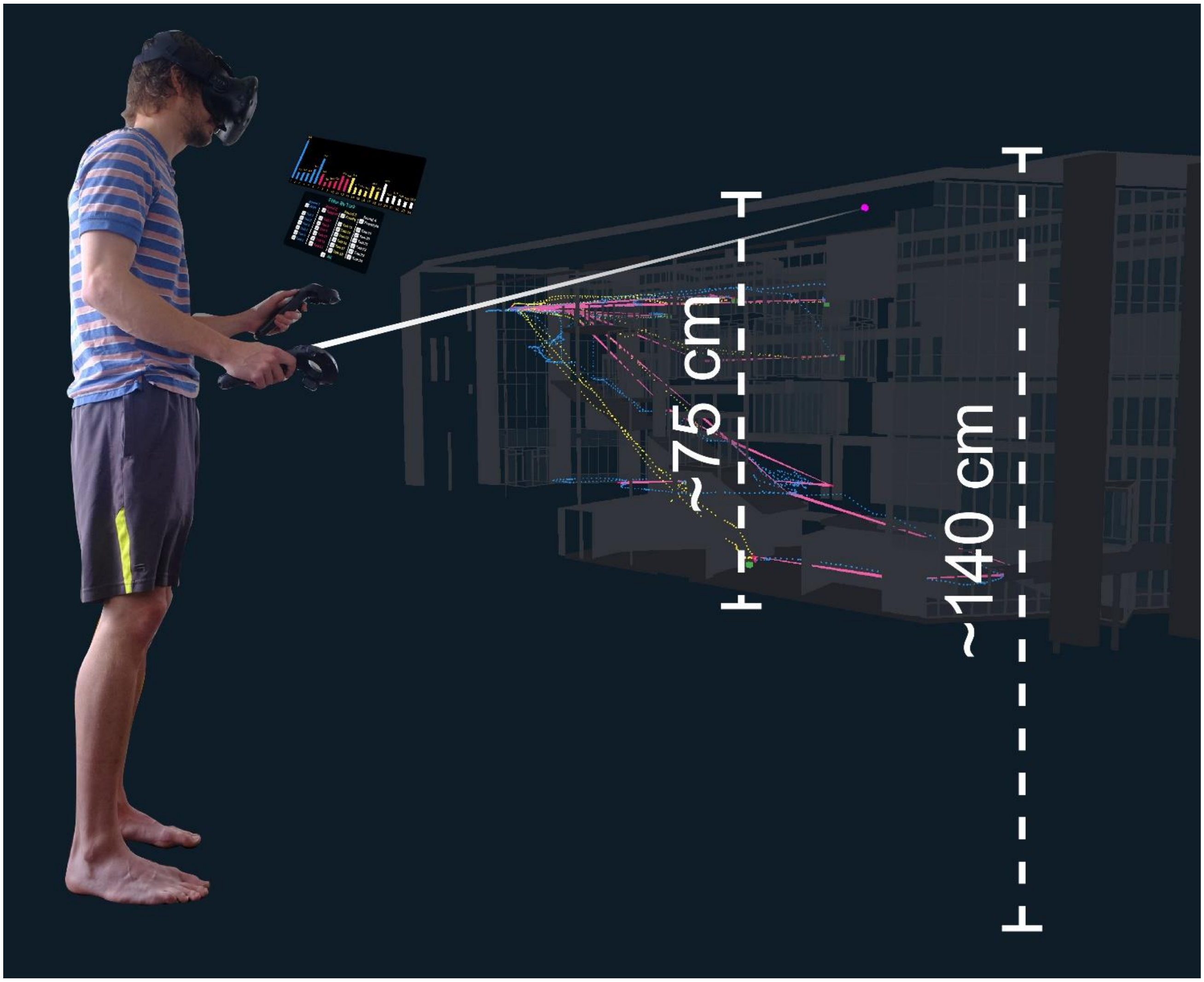

**Supplementary Figure 19.** Approximation of a user in relation to the 3D base map and visualization in VR.

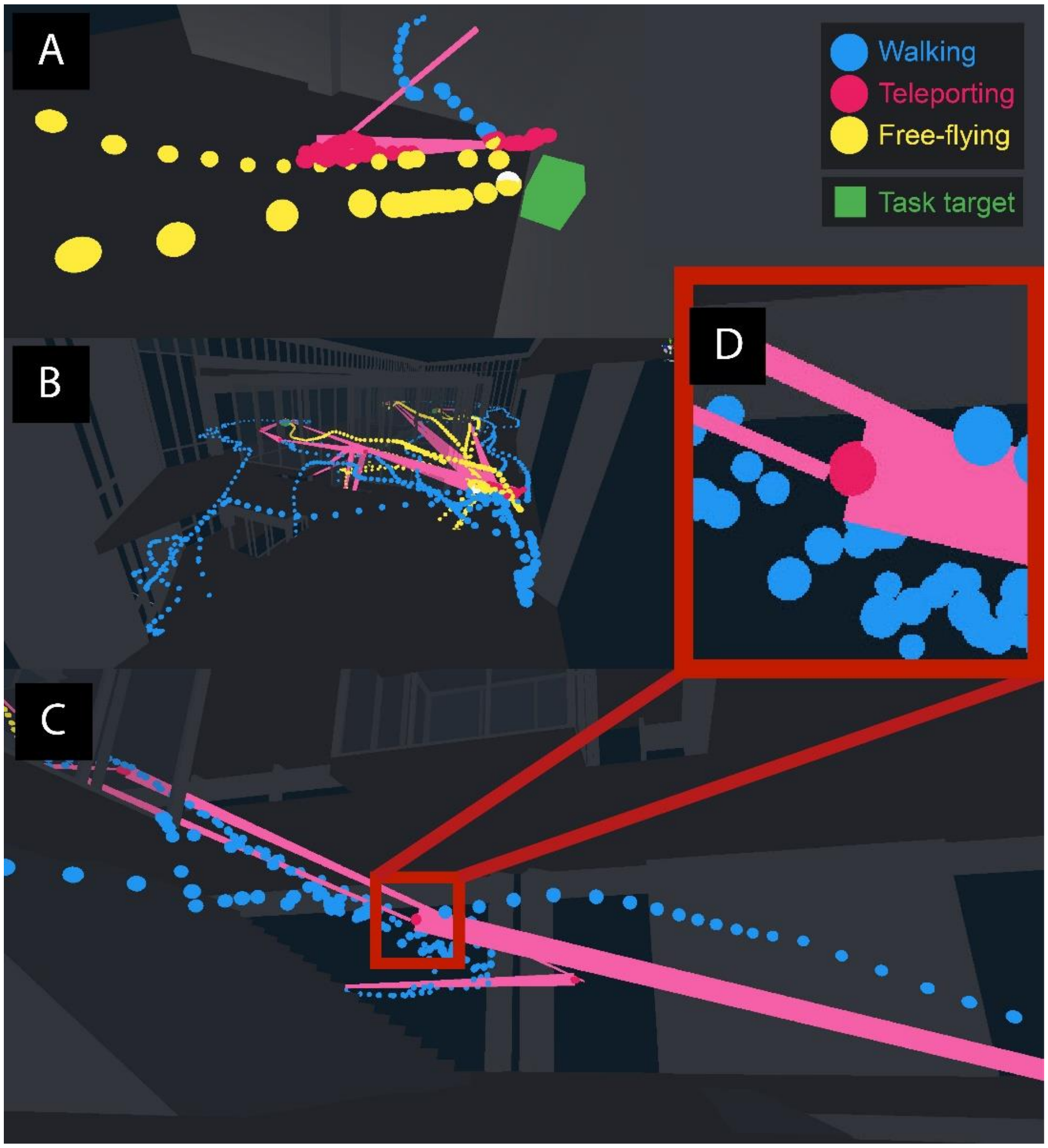

**Supplementary Figure 20.** Visual encoding details. A: a user's approaches to task #5. Note that there are two yellow trajectories, because they also used free-fly to finish this task when they had the choice. B: the start position on the fourth floor and the 24 trajectories leading away from it. C: a stop during a series of teleports. D: Note how the line leading to the teleport stop is thin, and the line leading away is wide, signaling the direction of the teleport.

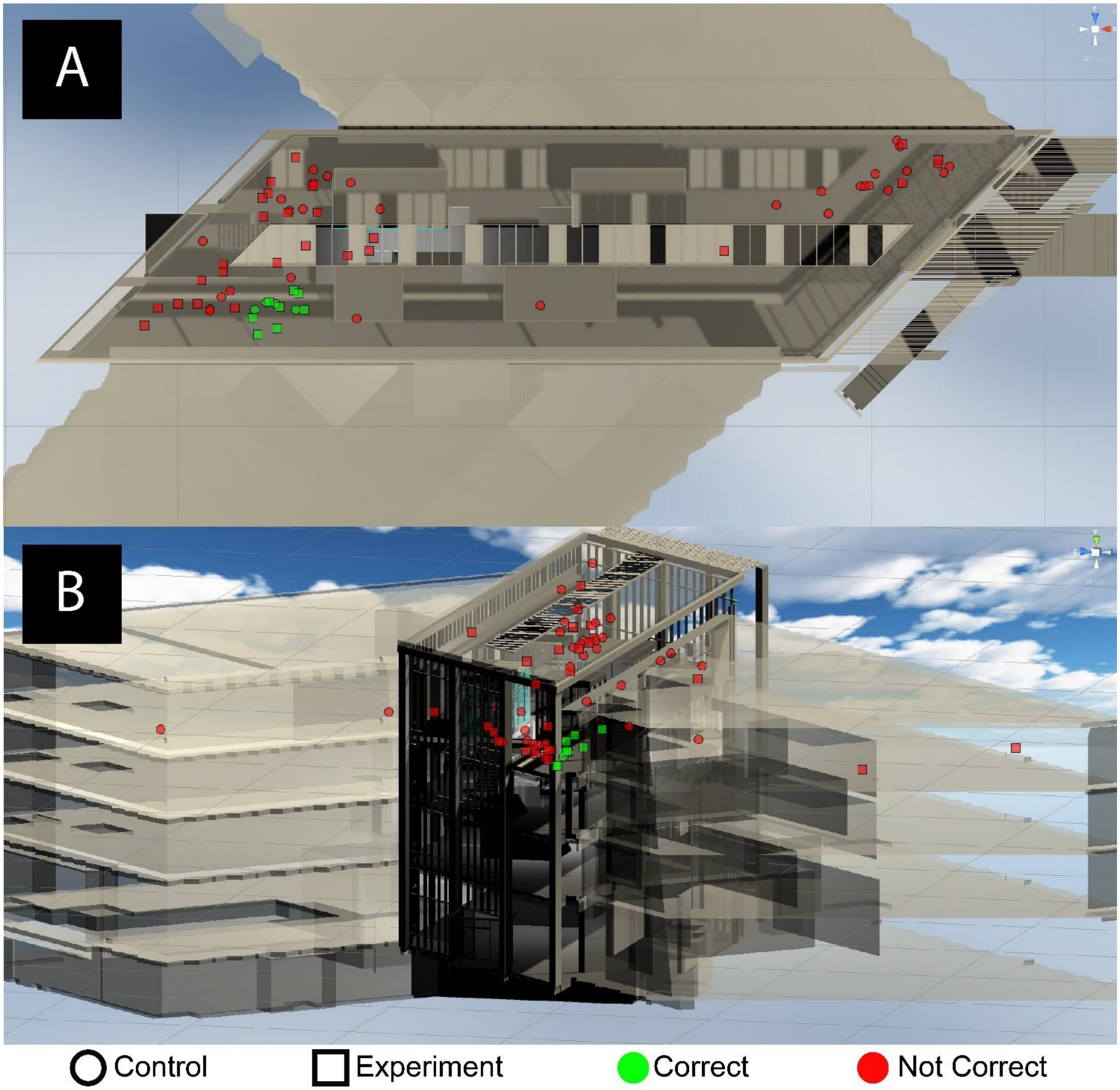

**Supplementary Figure 21.** Dot density maps pf click positions where users were asked to indicate where in these screenshots their start position was. A: top view. B: side view.

Like the rest of the mid-questionnaire, the visualization shows a stark difference between the cohorts. For the top view, there were 12 correct answers (4 for control, 8 for experiment); for the side view, there were only 10 correct answers (2 for control, 8 for experiment). The difference in correct answers for the side view between the cohorts for this particular task is significant (**$t$ = -2.0899, $p$ = 0.04158**). This is likely thanks to the Reflective Phase, because every trajectory in the Reflective Phase visualization marked the user's start position (see Supplementary Figure 20 B), thus giving an advantage to the experiment cohort.

While many users were at least able to determine that they had started somewhere in the atrium of Luddy Hall, a surprisingly large number of subjects mistook the barely modeled wings of the building for the location where they spent the experiment (see Supplementary Figure 21B). As becomes apparent from Supplementary Figure 21A, on the other, the top view helped users narrow down the amount of choices, and the majority of subjects picked one of the four corners of the atrium. Note that we found **no significant correlation** between how familiar users were with Luddy Hall as indicated on a 5-point Likert scale and the total score in the mid-questionnaire.

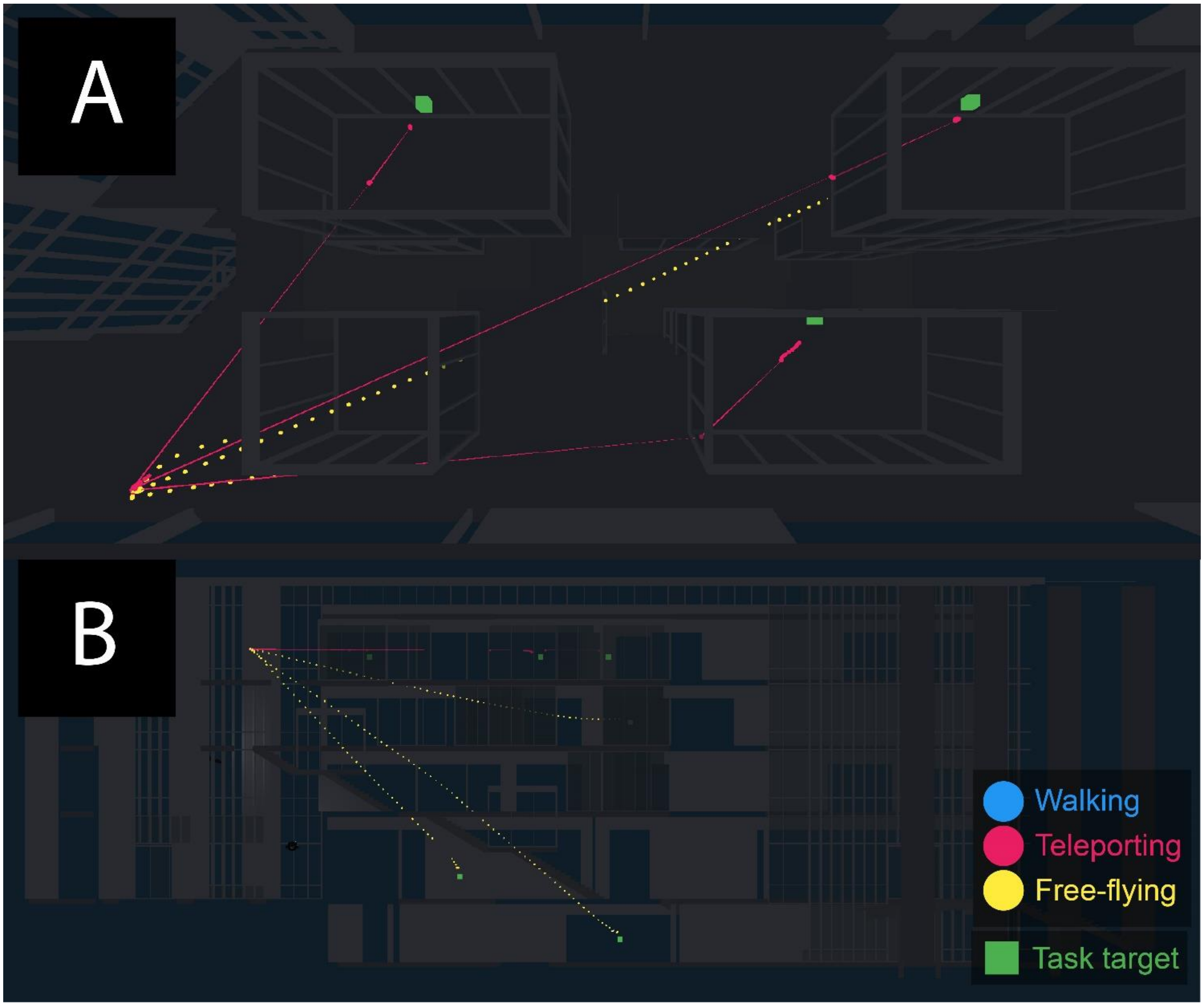

**Supplementary Figure 22.** The "winning strategy". A: using teleport for targets on the same floor as the start position. B: using free-fly for all others.

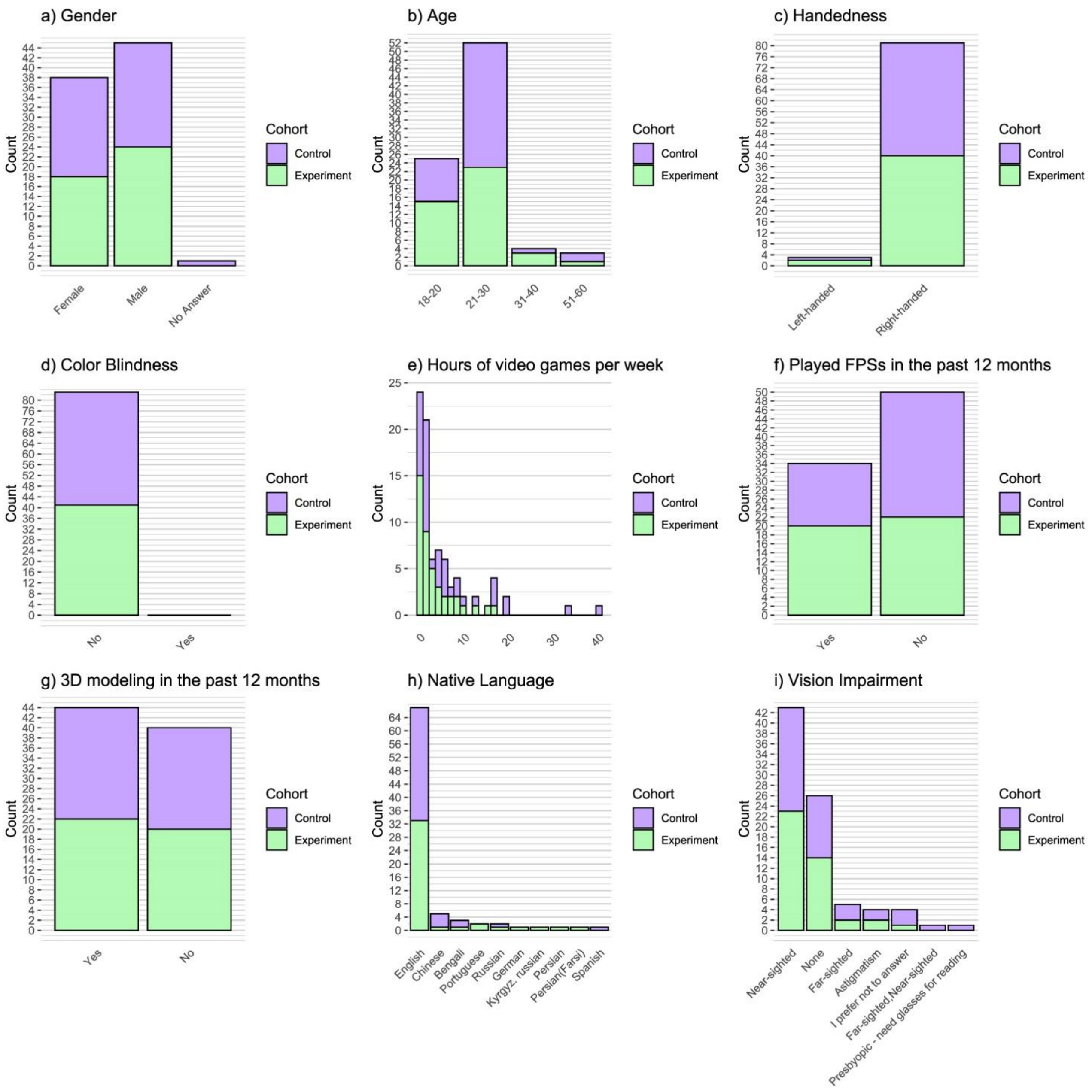

**Figure 23. Demographic make-up of RUI VR participants.**

## 5 Supplementary Tables

**Supplementary Table 1. Summary statistics of tool usage during Reflective phase.**

| | **VR Tabletop** | **VR Standup** |
|---|---|---|
| **Mean time spent** | 464.63 s (~7.7 mins) | 396.76 s (~6.6 mins) |
| **Mean distance traveled** | 26.88 m | 41.96 m (min = 13.635 m, max = 81.413) |
| **Mean total head rotation (y-axis)** | 7086.07 degrees (39.4 left-to-right head rotations) | 8434.43 degrees (46.9 left-to-right head rotations) |
| **Mean number of visible tasks** | 6.42 | 6.01 |
| **Times scrolled through length of dataset (H3a ✗)** | 5.15 | 2.68 |
| **Mean time spent without kidney (H3c ✓)** | 46.9% | 32.2% |

**Supplementary Table 2. Definition of variables for user behavior from the Reflective phase.**

| **Variable** | **Definition** |
|---|---|
| **total_time_spent** | time spent in the reflection phases |
| **distance_[INPUT DEVICE** | cumulated movement of left hand, right hand and head (in meters) |
| **degree_headrotationY** | Cumulated total degrees of head rotation around the y-axis |
| **head_upDownY** | Cumulated total head movement up and down the y-axis |
| **mean_rawSlider** | Average raw slider position ranging from 0 to 1 for each subject |
| **amountKidneyTurnoff** | Total number of times the kidney visualization was turned off |
| **time_without_kidney** | Share of reflective time spent with kidney turned off |
| **time_toggle_filter_usage** | Share of reflective time spent with other filter toggles used |
| **avg_task_visible** | Average of task numbers that were visible during reflective phase |
| **avg_number_tasks_visible** | Average numbers of tasks that were visible at the same time |

All the values reported below have been recorded in the **main** part of the Reflective phase.

**Supplementary Table 3. The p-values of the Kruskal-Wallis-Tests for left-handed/right-handed subjects and their performance (no significance). Note there were only 3 left-handed subjects. The column for VR Standup is empty, because there were only right-handed people in this setup.**

| | **Setup** | | |
|---|---|---|---|
| **Performance metric** | **2D Desktop** | **VR Tabletop** | **VR Standup** |
| **Completion time** | 0.142857142857143 | 0.428571428571429 | N/A |
| **Position accuracy** | 0.785714285714286 | 0.761904761904762 | N/A |
| **Rotation accuracy** | 1 | 0.64021164021164 | N/A |
| **Satisfaction** | 0.136345217441964 | 0.956806206691767 | N/A |